\documentclass[10pt,preprint]{aastex}

\usepackage{color}
\usepackage{ulem}

\shorttitle{Stellar Yields of Rotating First Stars. I.}
\shortauthors{Takahashi et al.}

\begin{document}

\title{Stellar Yields of Rotating First Stars. I. Yields of Weak Supernovae and Abundances of Carbon-enhanced Hyper Metal Poor Stars}

\author{
Koh Takahashi$^1$,
Hideyuki Umeda$^1$,
and
Takashi Yoshida$^2$
}
\affil{
$^1$ Department of Astronomy, The University of Tokyo.\\
$^2$ Yukawa Institute for Theoretical Physics, Kyoto University.\\
Accepted for publication in ApJ.}

\begin{abstract}

We perform stellar evolution simulation of first stars and calculate stellar yields from the first supernovae. The initial masses are taken from 12 to 140 M$_{\odot}$ to cover the whole range of core-collapse supernova progenitors, and stellar rotation is included, which results in efficient internal mixing. A weak explosion is assumed in supernova yield calculations, thus only outer distributed matter, which is not affected by the explosive nucleosynthesis, is ejected in the models. We show that the initial mass and the rotation affect the explosion yield. All the weak explosion models have abundances of [C/O] larger than unity. Stellar yields from massive progenitors of $>$ 40-60 M$_{\odot}$ show enhancement of Mg and Si. Rotating models yield abundant Na and Al. And Ca is synthesized in non-rotating heavy massive models of $>$ 80 M$_{\odot}$. We fit the stellar yields to the three most iron-deficient stars, and constrain the initial parameters of the mother progenitor stars. The abundance pattern in SMSS 0313-6708 is well explained by 50-80 M$_{\odot}$ non-rotating models, rotating 30-40 M$_{\odot}$ models well fit the abundance of HE 0107-5240, and both non-rotating and rotating 15-40 M$_{\odot}$ models explain HE 1327-2326. The presented analysis will be applicable to other carbon-enhanced hyper metal poor stars observed in the future. The abundance analyses will give valuable information about the characteristics about the first stars.

\end{abstract}

\keywords{nuclear reactions, nucleosynthesis, abundances -- stars: abundances -- stars: Population III -- stars: rotation}


\section{INTRODUCTION}

First stars, also known as Population III (Pop III) stars, are key drivers of the evolution of the early universe \citep[and references therein]{bromm&yoshida11}.
Ionizing photons emitted during the evolution of Pop III stars initiate the cosmic reionization.
Pop III stars are also the first nuclear reactors in the universe.
Pop III supernova explosions pollute ambient primordial gases with metals.
The first metal enrichment changes the cooling characteristics of primordial gases and initiates the shift of the stellar population from the metal-free Pop III stars to the metal-rich succeeding stellar populations.

Recent developments of theoretical calculations have revealed the characteristics of the formation of Pop III stars from the primordial gas clouds \citep[see][for reviews]{bromm13, glover13}.
Starting with the well determined initial structure at the cosmic recombination, ab-initio cosmological simulations identify the place where the first star formation takes place \citep[e.g.,][]{abel+02, bromm+02, yoshida+03}.
At the center of small dark matter halos of $10^5$-$10^6$ M$_{\odot}$ at the redshift of z $\sim$ 20-30, gravitationally unstable primordial gas clouds of $\sim$1000 M$_{\odot}$ start to collapse.
With the help of cooling by hydrogen molecules, the collapse continues until formation of a protostellar core of $\sim$0.01 M$_{\odot}$ \citep{yoshida+08}.
The evolution of the core during the following accretion phase remains unclear, because several processes, such as irradiation by the protostar \citep{mckee&tan08, hosokawa+11}, fragmentation of the accretion disk \citep{clark+11, greif+12}, and the magnetic field \citep{turk+12, machida&doi13} may significantly affect the result.
However, still robust information about the formed Pop III stars can be inferred from recent simulations.
\citet{hirano+14} calculated a lot of star formation gas clouds, and the result suggests that Pop III stars will show a wide initial mass function of $\sim$10-1000 M$_{\odot}$.
Also Pop III stars are suggested to have a fast rotation velocity at birth \citep{stacy+11b, stacy+13}.

Hence, demand for observational tests to check the theoretical estimates of the Pop III stellar characteristics has been increased.
And the requirement may be achieved by doing {\it abundance profiling} on metal poor stars.
Metal poor stars are traditionally used as a probe of the nucleosynthetic results of PopIII supernovae \citep[e.g.,][]{nomoto+13}.
This is because, the most metal poor stars may be second generation stars, which directly show the nucleosynthetic yield of mother Pop III stars.
Supernova explosions may pollute primordial gas inhomogeneously in the early universe, thus, stars formed from the unevenly polluted metal poor gases are expected to show characteristics of a single or a few supernova ejecta \citep{audouze&Silk95, tumlinson06}.

Motivated by these ideas, several surveys and observations have been carried out, increasing the number of observed metal poor stars (e.g., HK survey: \citealt{beers+92}; Hamburg/ESO survey: \citealt{christlieb03}; SEGUE survey: \citealt{yanny+09}, and for recent works, \citealt{yong+13, aoki+13}).
Among metal poor stars, stars showing smaller metallicity of -4 $\leq$ [Fe/H]\footnote{Here [A/B] = log$_{10}$($N_A/N_B$)$-$log$_{10}$($N_A/N_B$)$_{\odot}$, where the subscript ${\odot}$ denotes the solar value and $N_A$ and $N_B$ are the number abundances of elements A and B, respectively.} $<$ -3 are called extremely metal poor (EMP) stars, and similarly, stars of -5 $\leq$ [Fe/H] $<$ -4 and -6 $\leq$ [Fe/H] $<$ -5 are respectively called ultra metal poor (UMP) stars and hyper metal poor (HMP) stars \citep{beers&christlieb05}.
Owing to the large number of samples, some trends in their chemical abundances have been shown.
For example, EMP stars show decreasing [(Cr, Mn)/Fe] and increasing [(Co, Zn)/Fe] with decreasing [Fe/H] \citep[e.g.,][]{mcwilliam+95}.
Also, the large number of samples includes extremely rare, but a certain number of HMP stars.
Until now, three metal poor stars with [Fe/H] $<$ -5 are known.
They are HE 0107-5240 of [Fe/H] = -5.3 \citep{christlieb+02, christlieb+04}, HE 1327-2326 of [Fe/H] = -5.7 \citep{frebel+05, aoki+06}, and recently found SMSS 0313-6708 of [Fe/H] $<$ -7 \citep{keller+14}.
Peculiar characteristics of the abundance patterns are not only the low iron abundance, but also the enhancement of intermediate mass elements, such as carbon, nitrogen, oxygen, sodium, and magnesium.
The representative feature is a large abundance of carbon; they have [C/Fe] $\gtrsim$ -3.

It has been shown that a phenomenological model of a supernova explosion called the mixing-fallback model can naturally explain the abundance trends seen in EMP stars \citep{umeda&nomoto02, umeda&nomoto05, tominaga+07b, tominaga+14}.
In the model, an explosion is assumed not to eject the whole stellar mass, but some fraction of fully-mixed inner matter in addition to the outer matter is ejected.
Accordingly, this model has three basic parameters, the inner boundary of the mixing region corresponding to the initial mass of the compact remnant, $M_{cut}$, the outer boundary of the mixing region, $M_{mix}$, and the ejection fraction with which the matter in the mixing region escapes to interstellar space, $f$ \citep[see][]{tominaga+07b}.
Moreover, with variations in the explosion energy and the mixing-fallback parameters, abundance patterns of HMP stars are also well explained \citep{umeda&nomoto03, iwamoto+05, ishigaki+14}.
It has been shown that an explosion with a low iron yield in an iron poor environment is needed to explain observed large [C/Fe] ratios for HMP stars \citep[e.g., ][]{nomoto+13}.
Based on these successful explanations, two schematic illustrations are discussed to account for the mixing-fallback process.
The first one is a low energy supernova.
At the boundary region of a stellar core, deceleration by the reverse shock takes place \citep[e.g.,][]{hachisu+90, kifonidis+03}.
The deceleration accounts for the large fallback of inner matter, simultaneously explaining the mixing by the growth of the Rayleigh-Taylor instabilities.
The other one is a jet-like explosion \citep{tominaga+07a, tominaga09}, in which large fallback is attributed to the accretion of off-axis matter.

While previous works provided information about the mechanisms of supernova explosions, the progenitor characteristics such as initial masses and rotational velocities have not been constrained.
Firstly, a useful probe to constrain the initial mass of the progenitor star is still unknown.
This is because of the degeneracy in the abundance patterns and the progenitor masses \citep{umeda&nomoto05}.
In the mixing-fallback model, inner abundance ratios that account for heavy nuclei, such as iron-peak elements, are similar for more massive stars with more energetic explosions.
Secondly, a probe for the rotational velocity is unknown.
Previous calculations of Pop III progenitors only consider the case of non-rotating stars \citep{umeda&nomoto05, tominaga+07b, heger&woosley10}.
And there has been only a limited number of works for rotating Pop III evolution calculations \citep{marigo+03, ekstroem+08, yoon+12, chatzopoulos&wheeler12}, which do not compare their yields with observed abundance patterns.

The aim of this work is, thus, to obtain new knowledge of the first supernova yield abundances, which can be used to constrain the progenitor characteristics.
Calculated models have a wide initial mass range of 12-140 M$_{\odot}$, which is a likely mass range for core-collapse supernovae.
Additionally, in order to find a signature of stellar rotation, evolution of rotating stars is calculated.
So far, several works show that stellar rotation affects all results of stellar evolution \citep[e.g.,][]{meynet&maeder00, heger+00}, therefore rotation in Pop III stars is also expected to have an important consequence on the yields.
Indeed, it has been already shown that stellar rotation at small metallicities significantly affects the stellar nucleosynthesis, especially nitrogen production, which can explain a N/O plateau observed in metal poor host stars \citep{meynet&maeder02a, meynet&maeder02b, chiappini+06, hirschi07, meynet+10}.
Note that our calculation is limited to the case of single stellar evolutions.
Fragmentation during first star formation may result in a high fraction of binaries and/or multiple systems, and the binarity could affect the evolution and thus the nucleosynthesis results of the first stars \citep[e.g.,][]{stacy&bromm13}.
Effects of binarity are too complicated for our first investigation, but should be investigated in the future.

The paper is organized as follows.
Physical ingredients of the stellar evolution code and assumptions to calculate the yields are described in Section 2.
In Section 3, resulting features in the outer abundance that can be used to constrain the progenitor's initial parameters are summarized.
The best fit results to explain the observed abundance patterns of the three most iron-deficient stars of SMSS 0313-6708, HE 1327-2326, and HE 0107-5240, as well as model comparisons with previous works, are presented in Section 4.
Section 5 is dedicated to discussions, and summary and conclusion are presented in Section 6.


\section{METHOD}

\subsection{Code Description}

We use the latest version of the stellar evolution code described in \citet{takahashi+13}, \citet{umeda+12}, and \citet{yoshida&umeda11}.
Capability to treat stellar rotation is newly included in the code.
At first, basics of the evolution code, such as a nuclear reaction network and a mixing treatment are briefly presented.
Then, the treatment of stellar rotation is described.

\subsubsection{Basics of the Evolution Code}

The evolution code is used for calculating hydrostatic and hydrodynamic evolution of stellar structures.
The inertia term is included in later stages of evolution, after the oxygen burning phase for less massive stars (12-60 M$_{\odot}$) and after the helium burning phase for more massive stars (70-140 M$_{\odot}$).
Mass loss is almost neglected in a non-rotating calculation, and a very small rate of $\dot{M}(v_{rot}=0)=-10^{-14}$ M$_{\odot}$/yr is applied \citep{yoon+12}.
Chemical mixing is approximated by a diffusion equation, and the diffusion coefficient is calculated depending on several conditions including rotational effects.
In order to treat fast proton capture reactions (see Section 3.3.), the reaction network includes 260 isotopes from neutron to $^{80}$Br (see Table \ref{tab_isotope}).
The initial composition consists of $^1$H, $^2$H, $^3$He, $^4$He, and $^7$Li, and the initial primordial abundances are taken from \citet{steigman07}.
Adopted reaction rates are almost the same as in \citet{takahashi+13}, but 1.3 times as large as the value by \citet{caughlan&fowler88} is used for the rate of $^{12}$C($\alpha$, $\gamma$)$^{16}$O.

We assume convection appears in dynamically unstable regions of
\begin{eqnarray}
\nabla_\mathrm{rad} > \nabla_\mathrm{ad} + \frac{\varphi}{\delta}\nabla_\mu,
\end{eqnarray}
where $\varphi \equiv \frac{\partial \mathrm{ln} \rho}{\partial \mathrm{ln} \mu}$, $\delta \equiv -\frac{\partial \mathrm{ln} \rho}{\partial \mathrm{ln} T}$, and $\nabla_\mathrm{rad}$ and $\nabla_\mathrm{ad}$ are the radiative and adiabatic temperature gradients, respectively.
The temperature gradients in the convective regions are calculated by the mixing length theory (MLT) for the whole region of a star, and 2.0 is used for the mixing length parameter, $\alpha_{mix}$.
In order to treat convective mixing of chemical species, a diffusion coefficient by \citet{spruit92} is applied in the dynamically unstable regions, with the parameter of $f_{sc}$=0.3.
For the dynamically unstable regions, the mixing coefficient becomes large enough to account for the full-mixing.
The same coefficient is also applied to vibrationally unstable regions of
\begin{eqnarray}
\nabla_\mathrm{ad} + \frac{\varphi}{\delta}\nabla_\mu \ge \nabla_\mathrm{rad} > \nabla_\mathrm{ad},
\end{eqnarray}
in order to take into account the growth of the instability.
Additionally, overshooting of the convective motion at the edge of dynamically unstable regions is treated during core hydrogen and core helium burning phases.
An exponential decaying formula,
\begin{eqnarray}
D_{cv, ov} = D_{cv,0} \mathrm{exp}\Bigl( -2\frac{ \Delta r }{ f_{ov}H_{\mathrm{P},0} } \Bigl)
\end{eqnarray}
is applied, taking 0.02 as the parameter $f_{ov}$, where $D_{cv,0}$ and $H_{\mathrm{P},0}$ are the convective mixing coefficient and the pressure height at the edge of the convective region, and $\Delta r$ is a distance from the edge.

\subsubsection{Effects of Stellar Rotation}

Four effects of stellar rotation are taken into account in the code; deformation by the centrifugal force; angular momentum transfer in the star; matter mixing due to the development of rotational instabilities; and rotationally induced mass loss.
At first, a surface of constant pressure of $P$, $\psi_P$, is defined.
Then, shellular rotation is assumed as the rotation profile: the angular velocity is constant on the constant pressure surface \citep{zahn92}.
The mass coordinate of the calculation, $M_P$, is defined as the enclosed mass within the constant pressure surface, also the volume surrounded by the isobar, $V_P$, and the radius of the sphere which have the same volume as $V_P$, $4\pi r^3_P/3=V_P$, are defined.
Because of the deformation, the constant pressure surface is not a sphere, and the approximation for slow rotating cases \citep{denissenkov&vandenberg03} is adopted,
\begin{eqnarray}
r(\mathrm{cos}\theta) = a [1 - \epsilon \mathrm{P_2}(\mathrm{cos}\theta)],
\end{eqnarray}
where $\mathrm{P_2}(\mathrm{cos}\theta)$ is the second-degree of Legendre polynomial.
The scaling radius $a$ and the degree of rotation $\epsilon$ are related to $r_P$ as
\begin{eqnarray}
r_P &=& a(1 + \frac{3}{5}\epsilon^2 - \frac{2}{35}\epsilon^3)^{1/3} \\
\epsilon &=& \frac{\omega^2_P r_P^3}{3GM_P} \Bigl( \frac{a}{r_P} \Bigl)^3,
\end{eqnarray}
where $\omega_P$ is the angular velocity of the isobar, and $G$ is the gravitational constant, respectively.
Finally, an averaged quantity on the constant pressure surface is defined as
\begin{eqnarray}
\langle q \rangle = \frac{1}{S_P}\int_{\psi_P} q d\sigma,
\end{eqnarray}
where $S_P$ is a surface area of the isobar and $d\sigma$ is a element of the isobaric surface.

Basic equations of pressure and temperature gradients are modified to take into account the centrifugal force \citep{endal&sofia76, meynet&maeder97, heger+00},
\begin{eqnarray}
\frac{\partial \mathrm{log} P}{\partial \mathrm{log} M_P} &=& -\frac{GM^2_P}{4\pi r^4_P P} f_P - \frac{M_P}{4\pi r^2_P P} \Bigl( \frac{\partial^2 r_P}{\partial t^2} \Bigl)\\
\frac{\partial \mathrm{log} T}{\partial \mathrm{log} P} &=& \left \{
\begin{array}{l}
\nabla_\mathrm{MLT} \mbox{ (for convective regions)}\\
\nabla_\mathrm{rad}\frac{f_T}{f_P}[1+\frac{r^2_P}{GM_P f_P}(\frac{\partial^2 r_P}{\partial t^2})]^{-1} \mbox{ (for radiative regions)}
\end{array}
\right.
\end{eqnarray}
where $\nabla_\mathrm{MLT}$ is a temperature gradient determined by the MLT, and $\nabla_\mathrm{rad}=\frac{3\kappa}{16\pi acG}\frac{P L_P}{T^4 M_P}$ is a radiative temperature gradient.
$f_P$ and $f_T$ represent the modification by the centrifugal force,
\begin{eqnarray}
f_P &=& \frac{4\pi r^4_P}{G M_P S_P}\frac{1}{\langle g^{-1} \rangle} \\
f_T &=& \Bigl( \frac{4\pi r^2_P}{S_P} \Bigl)^2\frac{1}{\langle g \rangle \langle g^{-1} \rangle}.
\end{eqnarray}
where $g$ is the local effective gravity.

In a rotating star, several instabilities are assumed to develop.
Because matter mixing occurs due to these rotational instabilities, both chemical species and angular momentum are transported by the rotationally induced mixing.
Similar to chemical mixing, a diffusion approximation is adopted to the angular momentum transport \citep{endal&sofia78, pinsonneault+89, heger+00},
\begin{eqnarray}
\frac{\partial \omega_P}{\partial t}= \frac{1}{i} \Bigl( \frac{\partial}{\partial M_P} \Bigl) \Bigl[ (4\pi r^2_P \rho)^2 i \nu \Bigl( \frac{\partial \omega_P}{\partial M_P} \Bigl) \Bigl]
-\frac{\omega_P}{r_P} \Bigl( \frac{\partial r_P}{\partial t} \Bigl) \Bigl( \frac{\partial \mathrm{log}i}{\partial \mathrm{log} r_P} \Bigl),
\end{eqnarray}
where $i \equiv \langle g^{-1}r^2 \mathrm{sin}^2\theta \rangle / \langle g^{-1} \rangle$ is the specific moment of inertia and $\nu$ is the viscosity.
The first term represents the angular momentum transport by matter mixing, and the second term shows the local angular momentum conservation.
A precise treatment of mixing is so far difficult.
Hence, order-of-magnitude estimates are applied to determine both the viscosity and the diffusion coefficient.
Included hydrodynamical instabilities are the meridional (Eddington-Sweet) circulation, the dynamical and secular shear instabilities, the Solberg-H\o iland instability, and the Goldreich-Schubert-Fricke instability, and the corresponding viscosities, $D_{ES}, D_{DS}, D_{SS}, D_{SH}$, and $D_{GSF}$ are calculated following \citet{heger+00}.
In the calculation, the effect of the $\mu$-gradient is taken into account with the efficiency parameter $f_\mu$.
The value $f_\mu$=0.1 is adopted in this work \citep{brott+11a, yoon+12}.
For a magnetic instability, magnetic fields generated by the Spruit-Tayler dynamo are considered \citep{spruit02, heger+05}, and corresponding viscosity and diffusion coefficient, $\nu_{ST}$ and $D_{ST}$ are calculated.
Finally, in addition to mixing by convective instabilities, these values are summed up to determine total effective viscosity and diffusion coefficient,
\begin{eqnarray}
\nu &=& D_{cv} + D_{ES} + D_{DS} + D_{SS} + D_{SH} + D_{GSF} + \nu_{ST} \\
D &=& D_{cv} + f_c\times(D_{ES} + D_{DS} + D_{SS} + D_{SH} + D_{GSF}) + D_{ST}.
\end{eqnarray}
In the summation, a parameter $f_c$, which represents the difference of the efficiency between the viscosity and the diffusion coefficient by hydrodynamic instabilities in a rotating medium, is used.
The value $f_c$=1/30 is applied in this work, according to \citet{heger+00}.

The last effect of stellar rotation is the rotationally induced enhancement of mass loss \citep[$\Omega \Gamma$-limit,][]{Langer98, Maeder&Meynet00}.
According to \citet{yoon+10, yoon+12}, the enhancement of the mass loss rate is calculated as
\begin{eqnarray}
\dot{M} = -\mathrm{min}\Bigl[ |\dot{M}(v_{rot}=0)|\times \Bigl( 1-\frac{v_{rot}}{v_{crit}} \Bigl)^{-0.43}, 0.3\frac{M}{\tau_\mathrm{KH}} \Bigl],
\end{eqnarray}
where $v_{rot}$ and $v_{crit}=\sqrt{GM(1-\frac{L}{L_\mathrm{Edd}})/R}$ are the rotation velocity and the critical rotation velocity at the surface of the star, $M$, $R$, and $L$ are mass, radius, and luminosity of the star, $L_\mathrm{Edd}$ is the Eddington luminosity, and $\tau_\mathrm{KH}$ is the Kelvin-Helmholtz time scale, respectively.

\subsection{Initial Parameter Sets for Stellar Evolution Calculations}

Stellar evolution of 24 progenitor models is calculated for a wide initial parameter range.
Metallicity is set to be zero, and two initial parameters are used to specify the progenitor.
The first one is the initial mass of the model.
The initial mass range is from 12 M$_{\odot}$ to 140 M$_{\odot}$ so that the range for core-collapse supernovae can be covered by these models.
The next one is the initial rotation.
As for the initial rotation, we basically consider two cases: non-rotating and rotating models are presented.
To avoid complex discussions on chemically homogenous evolution from fast rotators \citep[e.g.,][]{yoon+12}, stars with moderate rotations of $v_{rot}/v_k$ $\sim$ 0.15 at ZAMS are calculated, where $v_k \equiv \sqrt{GM/R}$ is the Keplerian velocity at the surface.
In addition, more slowly rotating models of half- and quarter-rotation are calculated for 20, 30, and 40 M$_{\odot}$ cases.

Stellar evolution calculations are followed from deuterium burning phases until central densities reach $10^{10}$ g/cm$^3$ during the last collapse.
Calculated models are summarized in Table \ref{tab_models}.
The mass of the iron core, $M_\mathrm{Fe}$, is defined as the mass coordinate of the local peak of energy generation by silicon burning.
The CO core mass $M_\mathrm{CO}$, or mass of the base of a helium layer, is taken to be the mass coordinate at which the mass fraction of helium reaches 0.1.
Similarly, the top of the helium layer, $M_\mathrm{CO} + \Delta M_\mathrm{He}$, is defined as the mass coordinate where the mass fraction of hydrogen becomes 0.01.

Although there has been a lot of works on the structure of rotating stars, how to construct a proper model is still under debate.
Internal mixing in a rotating star will be the most influential physics for the evolution, but precise treatment of rotational mixing is difficult so far.
This is why we basically take only one rotating model for each mass in the calculation, except for the additional 20-40 M$_{\odot}$ models.
The presented rotating models show varieties of nucleosynthesis due to efficient internal mixing.
Our rotating models can thus be regarded as the representative results of efficient internal mixing by rotation.

\subsection{Assumptions on Supernova Explosions}

Stellar matter is somehow ejected by supernova explosions.
Since a precise treatment of the explosion is difficult, some assumptions are needed to estimate the stellar yield.
In this work, we assume a {\it weak explosion} for every yield calculation.
This means that, explosive nucleosynthesis by shock heating is too weak to change the abundance distribution in the progenitor star.
Secondly, we assume that the weak explosion only expels the stellar matter at the outer region of the star, which is loosely bound by gravity.
The first assumption will be appropriate especially for the outer region of a star.
For a calculation of explosive nucleosynthesis with a typical explosion energy of 10$^{51}$ erg, the outer abundance distribution from the carbon convective region is not much affected by shock heating.
Since only such unmodified outer materials are assumed to be ejected in this model, simple integration gives a consistent supernova yield, even though the explosive nucleosynthesis is not followed in the procedure.

With these two assumptions, ejected mass of an element $i$, $M_i$, is calculated as a function of $M_{in}$,
\begin{eqnarray}
M_i(M_{in}) = \int_{M_{in}}^\mathrm{M_\mathrm{surface}} X_i(M) dM,
\end{eqnarray}
where $M_{in}$ is the inner boundary of the ejection and $X_i$ is an abundance distribution of $i$ in terms of mass fraction.
In our model, material that distributes below $M_{in}$ is assumed to be completely captured by the central remnant, and does not contribute to the yield.
Compared with the mixing-fallback model, our results will agree with them, if the mixing-fallback parameters of outer boundary of the mixing region, $M_{mix}$, and the ejection fraction, $f$, are respectively specified to $M_{in}$ and zero.
Thus, mixing process during the explosion is not important ingredient in our model, while mixing process during the stellar evolution is of importance.
The inner boundary $M_{in}$ may relate to the explosion energy, however, the physical interpretation depends on what explosion mechanisms are assumed for the weak supernova.
We do not specify the explosion mechanism in this work, but discuss some possibilities in Section 5.1.

\subsection{Assumptions on Abundance Profiling}

In this work, results of the abundance profiling on HMP stars are presented.
The results are based on the assumption that the observed metal-deficient stars are the second generation stars, so that they should show the nucleosynthetic signatures of first generation mother stars.
Using the calculated $M_i$ and mass fraction of the element $i$ in the ISM, $X_{i,\mathrm{ISM}}$, resulting compositions of second generation stars, $X_{i,\mathrm{2nd}}$, can be written as
\begin{eqnarray}
X_{i,\mathrm{2nd}} &=& \frac{ M_i + X_{i,\mathrm{ISM}} M_\mathrm{ISM} }{ M_\mathrm{SN} + M_\mathrm{ISM} } \label{eq1}  \\
 &=& \frac{ M_i/M_\mathrm{SN} + X_{i,\mathrm{ISM}} D }{ 1 + D },
\end{eqnarray}
where $M_\mathrm{SN} = \sum_i M_i$ and $M_\mathrm{ISM}$ are total masses of the ejecta and the ISM, and $D=M_\mathrm{ISM}/M_\mathrm{SN}$ is the dilution factor.
For elements heavier than carbon, the elemental mass in the ISM, $X_{i,\mathrm{ISM}} M_\mathrm{ISM}$, is assumed to be much smaller than the ejected mass $M_i$, and thus negligible, since the ejection comes from the first supernova.
Under this assumption, compositions of second generation stars are calculated as
\begin{eqnarray}
X_{i,\mathrm{2nd}} = \frac{ M_i/M_\mathrm{SN} }{ 1 + D }
\end{eqnarray}
for elements heavier than carbon.
In order to compare the results with observations, solar standardized values of
\begin{eqnarray}
[i/j] = \mathrm{log} \Bigl( \frac{ X_{i,\mathrm{2nd}} }{ X_{j,\mathrm{2nd}} } \Bigl) - \mathrm{log} \Bigl( \frac{ X_{i, \odot} }{ X_{j, \odot} } \Bigl) \label{eq2}
\end{eqnarray}
are calculated with the solar values by \citet{asplund+09}.

In the abundance profiling, we basically consider the consistency of abundance patterns of intermediate-mass elements from carbon to silicon.
This is because the weak supernova models show varieties of those elements productions, and thus progenitor models can be constrained through the comparison.
On the other hand, the origin of heavy elements, such as iron-peak elements, is not uniquely determined by the model.
The discussion on the possible origins of those elements is given in Section 5.2.
Sometimes it is useful to compare an abundance pattern using iron as the base line, [$i$/Fe].
When we compare our results with observations using [$i$/Fe], a mass fraction of iron is taken as a free parameter.


\section{ABUNDANCE DISTRIBUTION IN PROGENITOR STARS}

Progenitor stars with different initial parameters have different conditions of density, temperature, and composition for nucleosynthesis.
Especially in a helium layer, which surrounds the inner core of oxygen and carbon, nucleosynthesis results in a variety of abundance distributions.
In the following subsections, how different conditions in each progenitor bring various nucleosynthetic results is presented for several elements.

\begin{figure}[htb]
		\centering
		\includegraphics[height=12cm, angle=270]{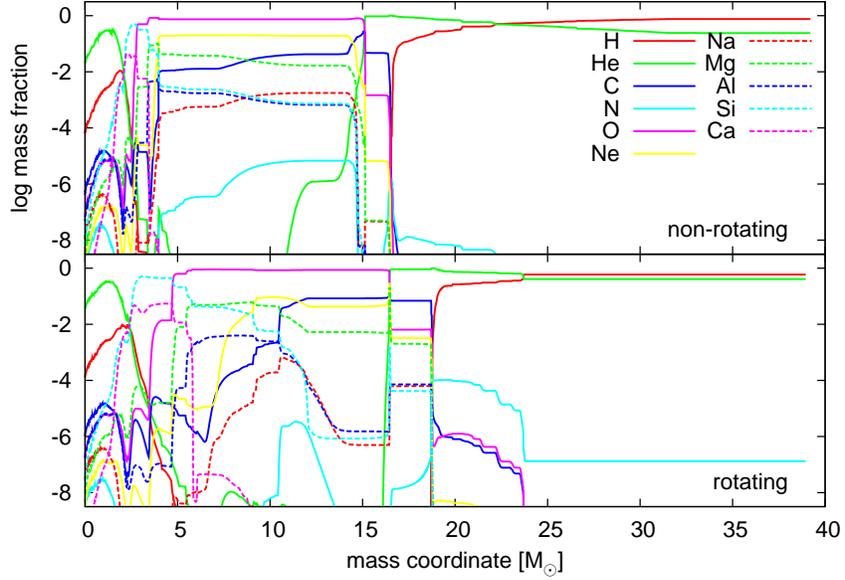}
	\caption{Mass fraction distributions of 40 M$_{\odot}$ models. The top panel corresponds to the non-rotating case, while the bottom to the rotating one. For the non-rotating model, $M_\mathrm{CO}$ and $\Delta M_\mathrm{He}$ are 15.07 M$_{\odot}$ and 1.571 M$_{\odot}$, while for the rotating model, these values become 16.45 M$_{\odot}$ and 2.387 M$_{\odot}$.}
	\label{dist_m40}
\end{figure}

\begin{figure}[htb]
		\centering
		\includegraphics[height=12cm, angle=270]{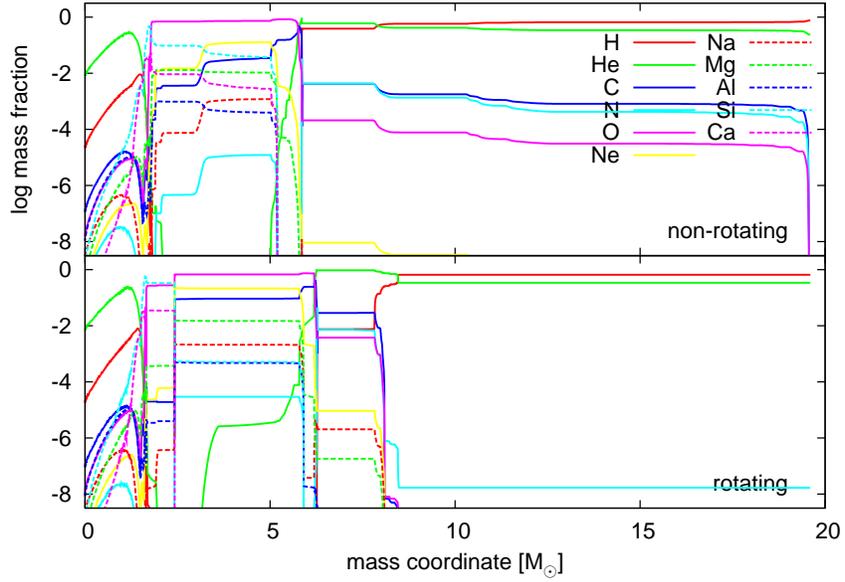}
	\caption{Same as Fig. \ref{dist_m20}, but for 20 M$_{\odot}$ models. For the non-rotating model, $M_\mathrm{CO}$ and $\Delta M_\mathrm{He}$ are 5.730 M$_{\odot}$ and 0.126 M$_{\odot}$, while for the rotating model, these values become 6.191 M$_{\odot}$ and 1.628 M$_{\odot}$. For the non-rotating model, a large mass fraction of hydrogen in the helium layer results from proton ingestion during core carbon burning phase.}
	\label{dist_m20}
\end{figure}

\begin{figure}[htb]
		\centering
		\includegraphics[height=12cm, angle=270]{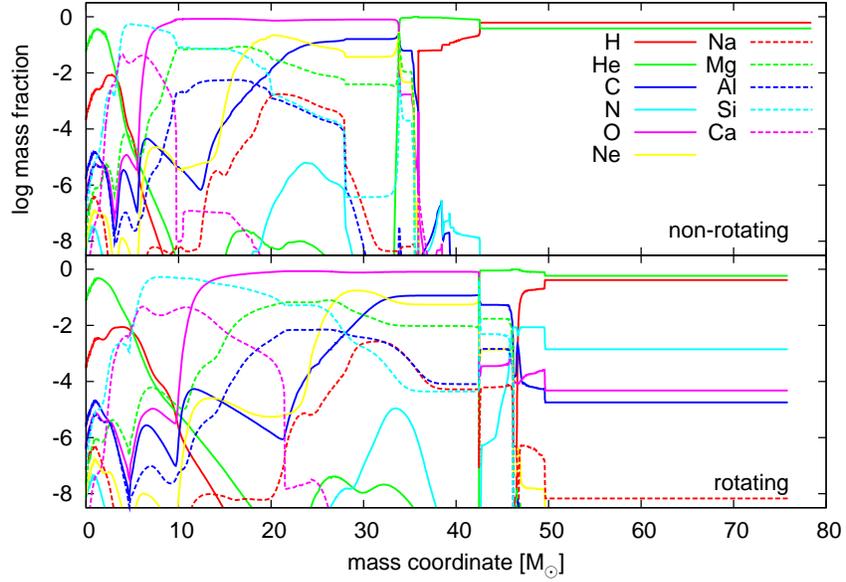}
	\caption{Same as Fig. \ref{dist_m40}, but for 80 M$_{\odot}$ models. For the non-rotating model, $M_\mathrm{CO}$ and $\Delta M_\mathrm{He}$ are 33.81 M$_{\odot}$ and 3.674 M$_{\odot}$, while for the rotating model, these values become 42.48 M$_{\odot}$ and 3.823 M$_{\odot}$.}
	\label{dist_m80}
\end{figure}

\begin{figure}[htb]
		\centering
		\includegraphics[height=12cm, angle=270]{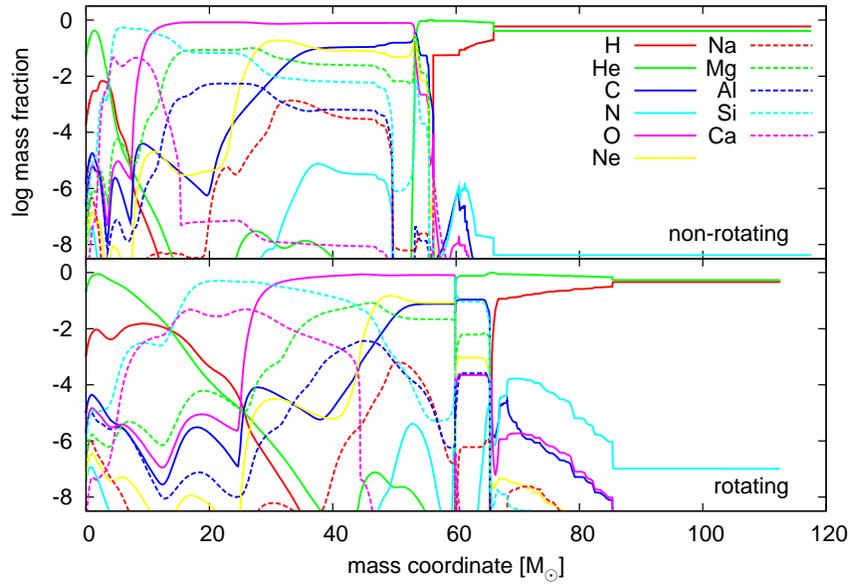}
	\caption{Same as Fig. \ref{dist_m40}, but for 120 M$_{\odot}$ models. For the non-rotating model, $M_\mathrm{CO}$ and $\Delta M_\mathrm{He}$ are 53.33 M$_{\odot}$ and 2.976 M$_{\odot}$, while for the rotating model, these values become 59.58 M$_{\odot}$ and 6.151 M$_{\odot}$.}
	\label{dist_m120}
\end{figure}

Figure \ref{dist_m40} shows mass fraction distributions for both non-rotating and rotating 40 M$_{\odot}$ models at the end of calculations.
In the helium layers, mass fractions of $^{12}$C and $^{16}$O do not show much differences.
On the other hand, intermediate mass elements of $^{23}$Na, $^{24}$Mg, and $^{27}$Al are well produced in the helium layer of the rotating model, and abundant $^{14}$N is distributed in the hydrogen envelope in the model.
We note that, while many isotopes are included in the reaction network for each element, the dominant isotope for intermediate-mass elements becomes the ordinal representative one, such as $^{12}$C, $^{14}$N, and $^{16}$O in most cases.
In these models, $M_\mathrm{CO}$ and $\Delta M_\mathrm{He}$ are 15.07 M$_{\odot}$ and 1.571 M$_{\odot}$ for the non-rotating model, and 16.45 M$_{\odot}$ and 2.387 M$_{\odot}$ for the rotating model, respectively.
Abundance distributions for other masses of 20, 80, and 120 M$_{\odot}$ are also presented in Figs. \ref{dist_m20}-\ref{dist_m120} to compare the different abundance distributions in the outer regions.

\subsection{Carbon and Oxygen}

\begin{figure}[htb]
		\centering
		\includegraphics[height=12cm, angle=270]{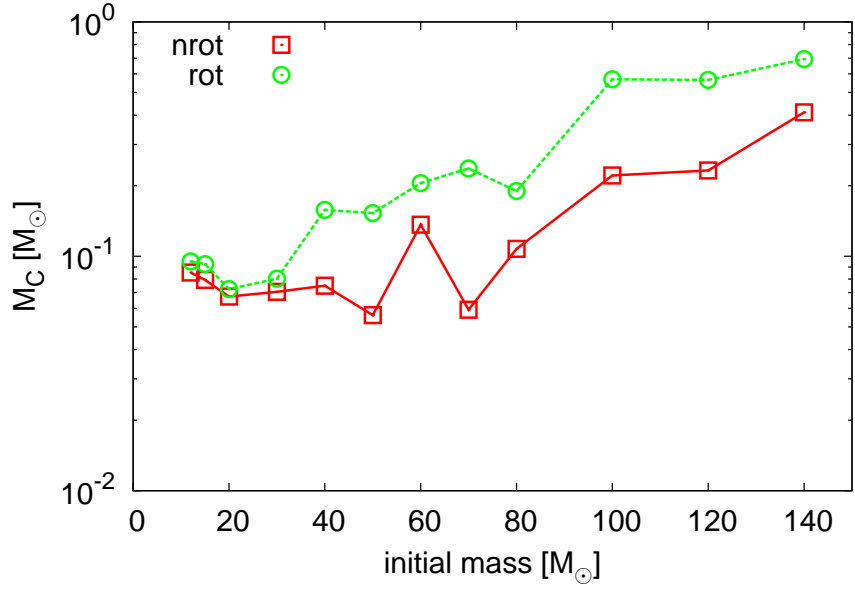}
	\caption{Integrated yield of carbon as a function of the initial mass. All isotopes of carbon are summed up. The range of integration is from the base of the helium layer to the surface. Results of non-rotating models are shown by red open squares connected by red solid lines, while green open circles with dashed lines correspond to rotating models.}
	\label{yield_c}
\end{figure}

\begin{figure}[htb]
		\centering
		\includegraphics[height=12cm, angle=270]{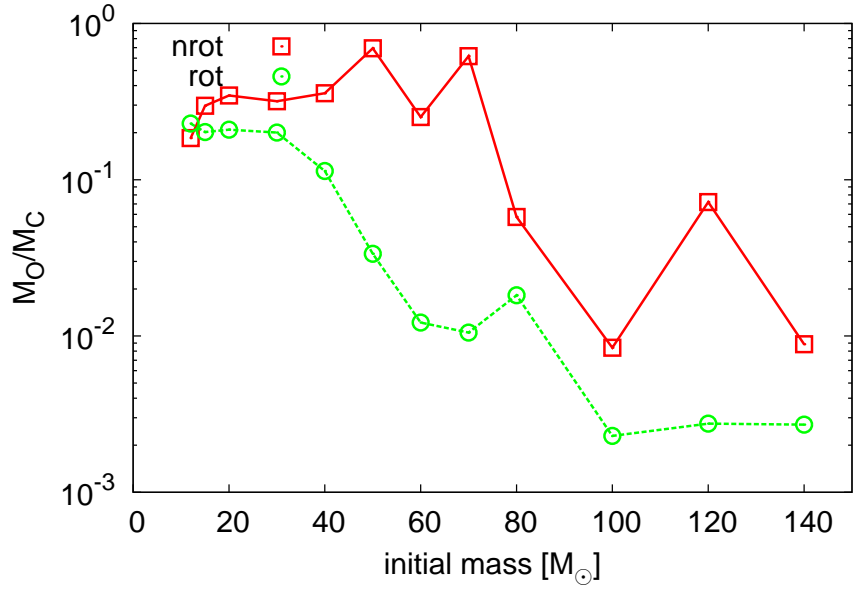}
	\caption{The production ratio between oxygen and carbon, $M_\mathrm{O}$/$M_\mathrm{C}$, as a function of initial mass. All isotopes of oxygen and carbon are summed up, respectively. Red open squares show non-rotating results, and green open circles with dashed lines show rotating results, respectively.}
	\label{ratio_o}
\end{figure}

Since helium is the most abundant element in a helium layer, the triple alpha reaction becomes the most influential reaction in the region.
This results in $^{12}$C production.
In the presence of both helium and carbon, alpha capture reaction onto $^{12}$C successively occurs, producing $^{16}$O.
Figure \ref{yield_c} shows $M_\mathrm{C}(M_{in}=M_\mathrm{CO})$ as the function of the initial mass, taking CO core masses as $M_{in}$ for each model, and mass ratio of $M_\mathrm{O}$/$M_\mathrm{C}$ is presented in Fig. \ref{ratio_o}.
In integration, all isotopes are summed up, while the most abundant isotopes are $^{12}$C and $^{16}$O.
Hereafter, all $M_{in}$ in a figure are fixed to be CO core masses, in order to show differences in outer abundance distributions among the models.

Production of carbon and oxygen also takes place at the center of the star during the core helium burning stage.
However, the resulting production ratio of O/C differs significantly between the two sites.
In the case of core helium burning, the resulting O/C ratio always exceeds unity.
On the other hand, a helium shell burning has a much smaller O/C ratio, since a large portion of helium remains during the evolution.

Three important characteristics on production of carbon and oxygen in a helium layer can be inferred from Figs. \ref{yield_c} and \ref{ratio_o}.
Firstly, production of carbon and oxygen takes place in all of the models.
This is because the temperature at the helium burning shell is high enough to allow the triple alpha and alpha capture reactions to occur.
Secondly, the O/C ratio does not exceed unity in all models.
And finally, the heavier the initial mass of the progenitor is, the smaller the resulting O/C ratio is.
Carbon and oxygen production with a small O/C can be regarded as a general nucleosynthetic signature of elemental production in helium layers, and more massive stars will have a smaller O/C ratio.

\subsection{Magnesium and Silicon}

Reflecting the different base temperature of helium layers, intermediate mass alpha elements such as magnesium and silicon show initial mass dependence.
These alpha elements are produced via a series of alpha capture reactions.
Because the reaction speed of an alpha capture is slower for a heavier element, higher temperature is needed to synthesize heavier alpha elements.
Initial mass of the model is the most influential physical parameter to determine the temperature.
Therefore, heavier alpha elements are only synthesized in more massive stars.

\begin{figure}[htb]
		\centering
		\includegraphics[height=12cm, angle=270]{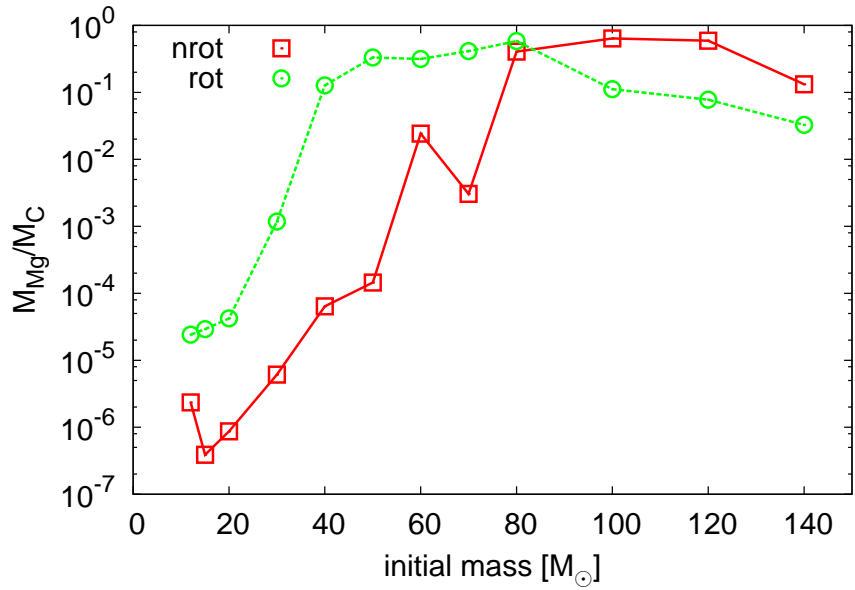}
	\caption{Same as Fig. \ref{ratio_o}, but for magnesium.}
	\label{ratio_mg}
\end{figure}

\begin{figure}[htb]
		\centering
		\includegraphics[height=12cm, angle=270]{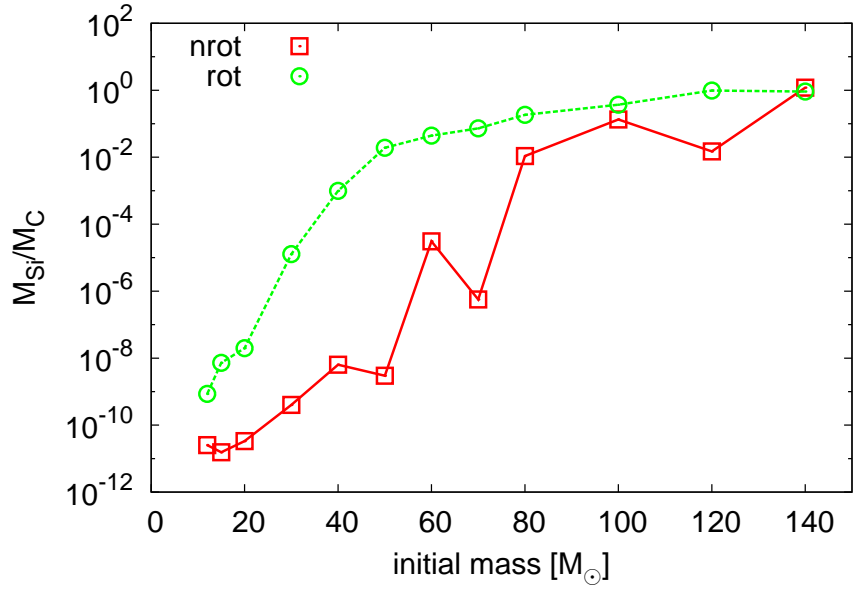}
	\caption{Same as Fig. \ref{ratio_o}, but for silicon.}
	\label{ratio_si}
\end{figure}

As for magnesium production, efficient production occurs for rotating $\ge$ 40 M$_{\odot}$ models and non-rotating $\ge$ 60-80 M$_{\odot}$ models.
The production ratio between magnesium and carbon is shown in Figure \ref{ratio_mg}.
The initial mass dependence can be characterized by the steep rise in less massive models and the plateau in more massive models.
The magnesium is mainly produced as $^{24}$Mg through $^{20}$Ne($\alpha$, $\gamma$)$^{24}$Mg.
The increase of the magnesium yield is a result of more efficient alpha capture reactions in massive models, and the plateau is due to the consumption of seed elements of $^{20}$Ne.
Silicon is produced by $^{24}$Mg($\alpha$, $\gamma$)$^{28}$Si, and the same trend on the progenitor initial mass is also found in the silicon production ratio, shown in Fig. \ref{ratio_si}.
Based on the trend in the initial mass dependence, the pattern of the produced alpha elements can be used as a probe of the progenitor's initial mass.

\subsection{Calcium}

\begin{figure}[htb]
		\centering
		\includegraphics[height=12cm, angle=270]{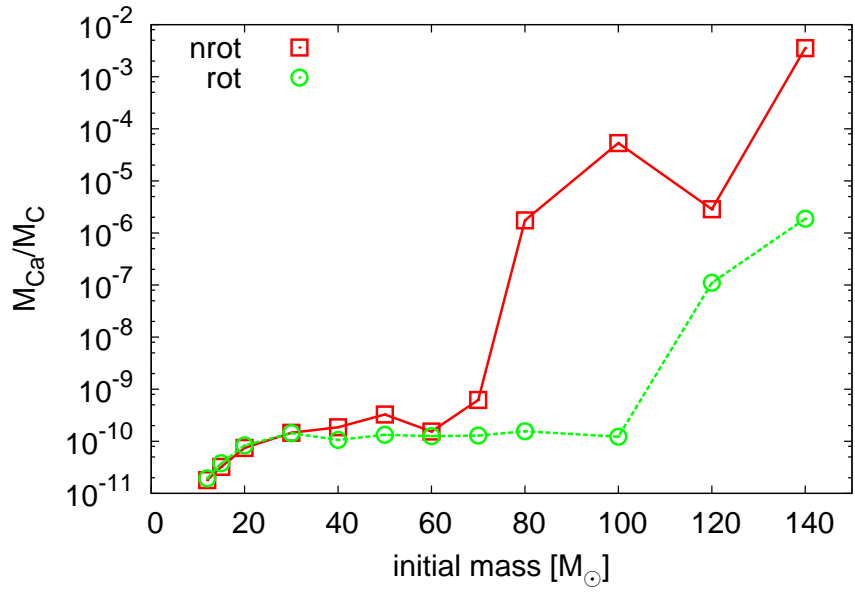}
	\caption{Same as Fig. \ref{ratio_o}, but for calcium.}
	\label{ratio_ca}
\end{figure}

Figure \ref{ratio_ca} shows the production ratio of calcium.
For rotating models, only the two most massive models of 120 M$_{\odot}$ and 140 M$_{\odot}$ show the enhancement.
It is fast alpha capture reactions at the base of the helium layer that synthesize calcium in these rotating models.
In this small region, other alpha elements of $^{28}$Si, $^{32}$S, and $^{36}$Ar are also produced, and by ($\alpha$, p) reactions on these alpha elements, some odd species of $^{31}$P, $^{35}$Cl, and $^{39}$K are synthesized as well.
For non-rotating models, abundant calcium production occurs for stars of $\ge$ 80 M$_{\odot}$.
Interestingly, a totally different nuclear process accounts for the production in non-rotating cases.

\begin{figure}[htb]
		\centering
		\includegraphics[height=12cm, angle=270]{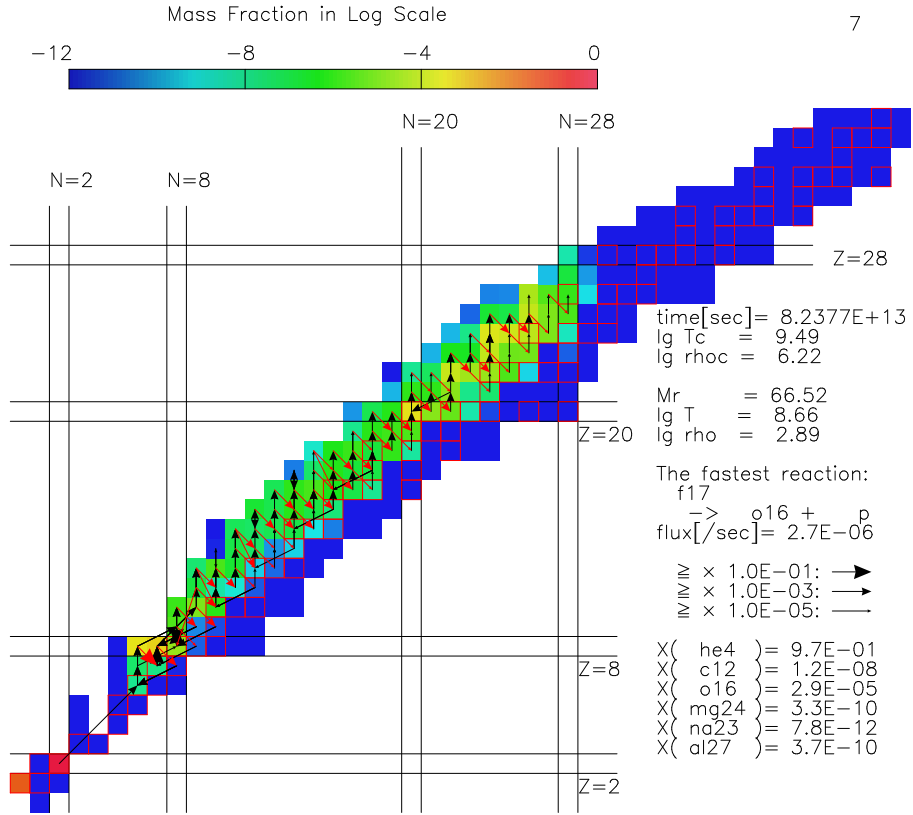}
	\caption{A nuclear chart with arrows showing fast reactions at a particular time, $t \sim 8.24 \times 10^{13}$ sec. Reactions at the base of the hydrogen burning shell of the non-rotating 140 M$_{\odot}$ are shown. Presented boxes correspond to different isotopes included in the reaction network, x- and y-axis show neutron and proton numbers respectively, colors show the mass fraction of each isotope, and red squares are for stable isotopes. Three different sizes of arrows show different magnitudes of fluxes normalized by the fastest reaction. Black arrows correspond to thermonuclear reactions, while red arrows correspond to reactions involving weak interactions.}
	\label{fig_pcap}
\end{figure}

The calcium production in non-rotating models is attributed to proton capture reactions in a hydrogen burning shell.
Similar to helium shell burning, the temperature of the hydrogen burning shell increases as the core of the star contracts.
If the base temperature gets high enough, break-out reactions from the CNO cycle take place \citep{wiescher+99}.
These reactions occur at the base of the hydrogen envelope of non-rotating $\ge$ 80 M$_{\odot}$ models in our calculation, resulting in production of proton rich isotopes including $^{40}$Ca.
Figure \ref{fig_pcap} shows how the reaction goes in the non-rotating 140 M$_{\odot}$ model after the central carbon burning phase.
Since the model has the largest initial mass in all of our models, the base temperature in the hydrogen burning region is very high, log $T_{base, \mathrm{H}}$ = 8.66 at this phase.
Accordingly, very efficient proton capture reactions take place in the region.
This calcium synthesis does not work in a rotating model, since the hydrogen envelope does not reach a high enough temperature for the break out to occur (see the next subsection).

\subsection{Nitrogen}

In massive stars, nitrogen is synthesized by the CNO cycle.
Since the reaction requires seed elements of carbon or oxygen in advance, not much nitrogen is synthesized in Pop III massive stars during core hydrogen burning phase.
The hydrogen burning shell does not have seed elements for canonical non-rotating models, thus these models do not produce much nitrogen during the evolution.
On the other hand, rotationally induced mixing accounts for the matter transportation from the helium burning core to the hydrogen burning shell in the case of rotating models.
As a consequence of the transportation of seed materials, the CNO cycle takes place in the hydrogen burning shell and all rotating models produce nitrogen during the core helium burning phase \citep{meynet&maeder02a, meynet&maeder02b}.

Nitrogen firstly distributes both in the hydrogen envelope and in the outer region of the helium core.
For nitrogen in the hydrogen envelope, matter mixing accounts for the enhancement.
The mixing processes are rotation induced mixing for less massive stars of $\leq$30 M$_{\odot}$, and convective mixing in a small convective region that appears in the early core helium burning phase for massive stars of $\geq$40 M$_{\odot}$.
These mixing processes transport nitrogen enriched material from the base of the hydrogen burning shell to the hydrogen envelope.
For nitrogen in the outer region of the helium core, matter accretion onto the helium core accounts for the nitrogen enrichment.
During the core helium burning phase, hydrogen shell burning increases the mass of the helium core.
The accreting matter has a large abundance of nitrogen, and all rotating models form nitrogen-rich helium layers at the end of the core helium burning phase.
Nitrogen in the helium layer accounts for the nitrogen yield for less massive models of $\leq$20 M$_{\odot}$.
On the other hand, for massive models of $\geq$30 M$_{\odot}$, most nitrogen in the helium layer is converted into $^{22}$Ne in later evolution stages (see the next subsection.), and do not contribute to the nitrogen yield.

\begin{figure}[htb]
		\centering
		\includegraphics[height=12cm, angle=270]{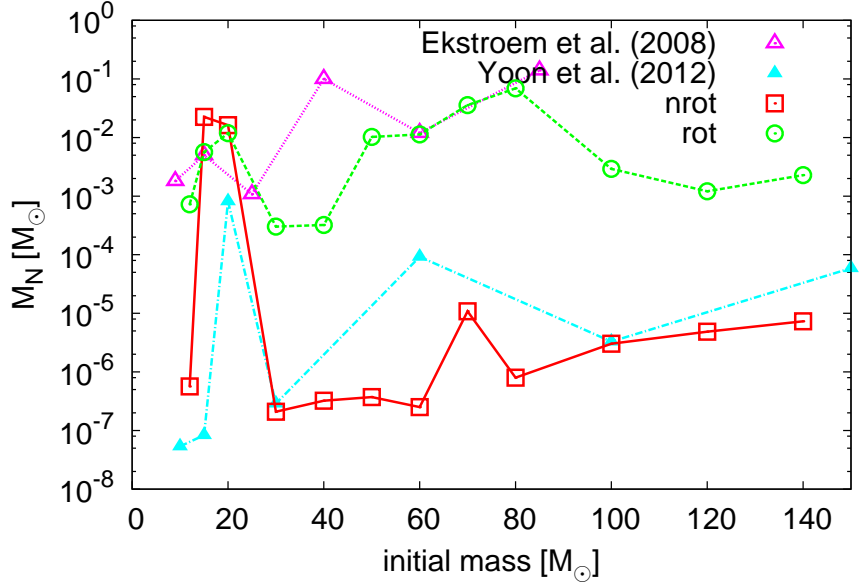}
	\caption{Same as Fig. \ref{yield_c}, but for nitrogen. In addition to our results shown by red squares (non-rotating models) and green circles (rotating models), results of rotating models from previous works are plotted. Magenta-open triangles show results by \citet{ekstroem+08} and blue-filled triangles are results of models of $v_{ini}/v_{k}$ = 0.2 by \citet{yoon+12}.}
	\label{ratio_n}
\end{figure}

Figure \ref{ratio_n} shows yields of nitrogen, $M_\mathrm{N}$, by several stellar calculations.
Yield data other than our results are rotating models of \citet{ekstroem+08} and $v_{ini}/v_{k}$ = 0.2 models of \citet{yoon+12}.
Since almost all of the nitrogen is distributed in hydrogen and helium layers at the last stage of the evolution, the graph shows the total yields of nitrogen in the calculations.
Our rotating models and those by \citet{ekstroem+08} show the enhancement of nitrogen.
Not much enhancement is seen in the models by \citet{yoon+12}, suggesting the uncertainty of treatment of stellar rotation.

\begin{figure}[htb]
		\centering
		\includegraphics[height=12cm, angle=270]{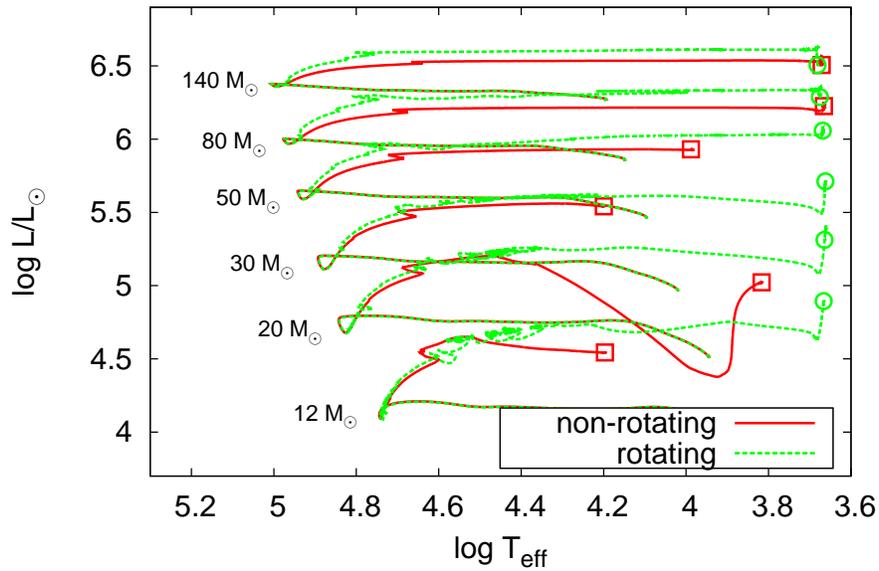}
	\caption{HR diagram of 12-140 M$_{\odot}$, non-rotating (red-solid lines) and rotating (green-dushed lines) models. Red squares show the end points for non-rotating models, and green circles show those of rotating models.}
	\label{hr_diagram}
\end{figure}

The active CNO cycle affects not only chemical distribution but also the hydrostatic structure of the progenitor star.
Hydrogen burning by the CNO cycle proceeds much faster than hydrogen burning by pp-chain.
Hence, a total amount of processed hydrogen significantly increases, and this results in a much thicker helium shell in a rotating star compared with a non-rotating counterpart.
Accordingly, the base of a helium shell is in a deeper region of the star, and the base temperature of the helium burning shell is higher for rotating models.
On the other hand, the base temperature of the hydrogen burning shell is lower for rotating stars.
This is due to an envelope inflation by the intense energy generated by the CNO cycle, by which every rotating model becomes a red-giant.
Only massive models of $>$ 70 M$_{\odot}$ become red-giants in non-rotating cases (see Fig. \ref{hr_diagram}).

We note that nitrogen production can be seen in some models with initial masses of 15-20 M$_{\odot}$ even for non-rotating cases.
This nitrogen production is due to hydrogen ingestion into the convective helium burning shell that occurs at a later evolutionary stage.
While the phenomenon causes some important changes in both stellar structures and chemical distributions, it may require some further attentions for the calculation.
We later discuss the difficulty and uncertainty to accurately treat the physics in Section 5.3.

\subsection{Sodium and Aluminum}

\begin{figure}[htb]
		\centering
		\includegraphics[height=12cm, angle=270]{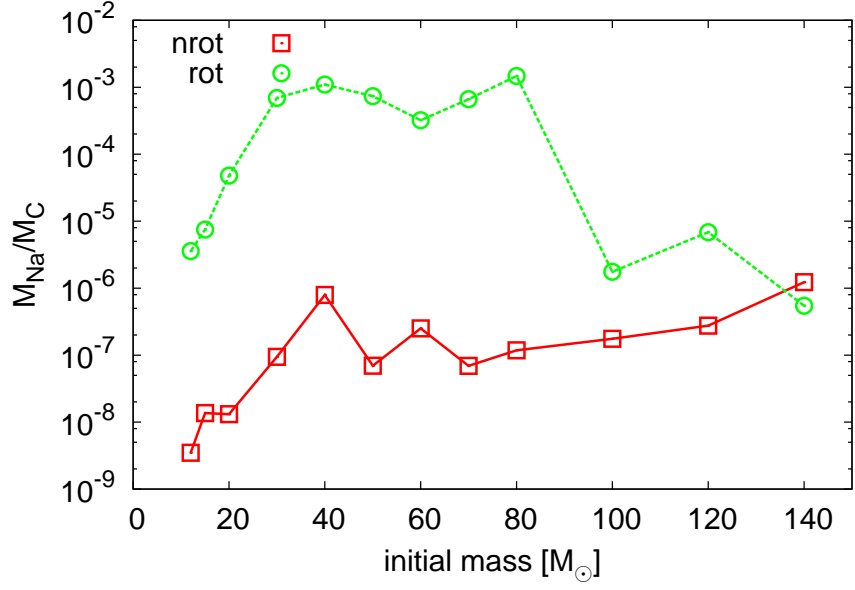}
	\caption{Same as Fig. \ref{ratio_o}, but for sodium.}
	\label{ratio_na}
\end{figure}

\begin{figure}[htb]
		\centering
		\includegraphics[height=12cm, angle=270]{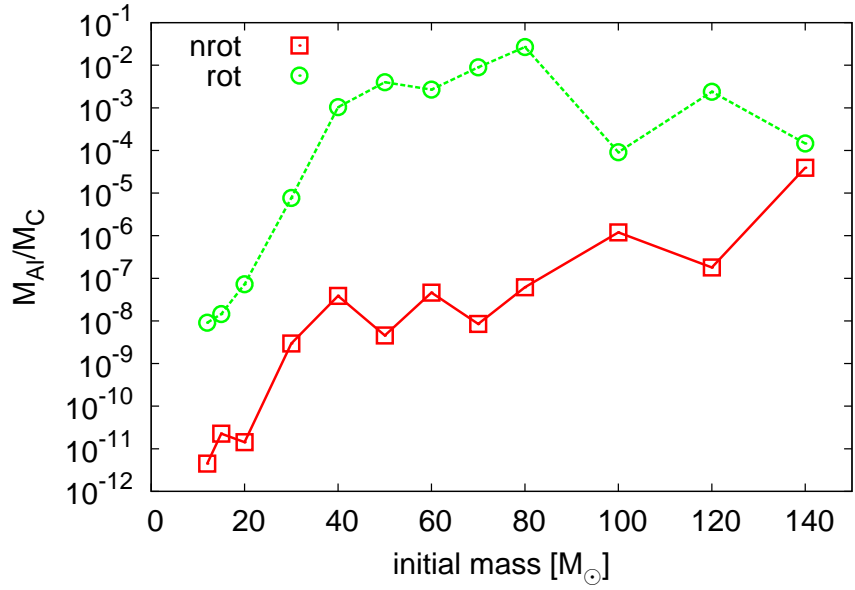}
	\caption{Same as Fig. \ref{ratio_o}, but for aluminum.}
	\label{ratio_al}
\end{figure}

The production ratios of sodium and aluminum are shown in Figs. \ref{ratio_na} and \ref{ratio_al}.
Rotating models show clear enhancement in production of these odd nuclei.
This is due to the existence of $^{14}$N in a helium layer in rotating models.
Because of the shell helium burning, abundant nitrogen in the region results in free neutron emission as a consequence of a series of alpha captures.
At first, alpha capture reactions produce $^{22}$Ne from $^{14}$N,
\begin{center}
$^{14}$N($\alpha, \gamma$)$^{18}$F($\beta^+ \nu_\mathrm{e}$)$^{18}$O($\alpha, \gamma$)$^{22}$Ne.
\end{center}
Then a free neutron is emitted by another alpha capture by $^{22}$Ne,
\begin{center}
$^{22}$Ne($\alpha,$ n)$^{25}$Mg.
\end{center}
When the nitrogen is absorbed by seed elements of $^{22}$Ne, $^{25}$Mg, and $^{26}$Mg, both $^{23}$Na and $^{27}$Al are produced as a result of the successive beta decays,
\begin{center}
$^{22}$Ne(n, $\gamma$)$^{23}$Ne($\beta^- \bar{\nu}_\mathrm{e}$)$^{23}$Na, \\
($^{25}$Mg(n, $\gamma$)) $^{26}$Mg(n, $\gamma$)$^{27}$Mg($\beta^- \bar{\nu}_\mathrm{e}$)$^{27}$Al.
\end{center}
Sodium and aluminum productions also occur in a hydrogen burning shell of rotating models.
These are due to the Ne-Na and the Mg-Al chains, since tiny fractions of neon and magnesium are transported from the base of the helium layer to the hydrogen burning shell by rotationally induced mixing.
However, sodium and aluminum production by these proton capture reactions is much less effective than the n-capture processes explained above.

Both the production ratios of sodium and aluminum have very similar dependence on the initial mass.
For less massive stars of $\lesssim$ 40 M$_{\odot}$, these production ratios show steep increase with the initial mass.
There is a plateau in the range from 40 M$_{\odot}$ to 80 M$_{\odot}$.
And these odd elements are less produced in the most massive stars of $\gtrsim$ 100 M$_{\odot}$.
These trends in the initial mass are due to the temperature dependence of related nuclear reactions.
The lower production in less massive stars can be understood as follows:
The alpha capture by $^{22}$Ne requires a temperature higher than $\sim$10$^{8.6}$ K, thus less massive stars do not have a sufficiently large flux of neutrons.
In addition, less massive stars do not have sufficiently abundant $^{24}$Mg as the seed element of $^{25}$Mg, since the alpha capture by $^{20}$Ne requires a temperature higher than $\sim$10$^{8.7}$ K.
This reduces aluminum production. 
On the other hand, lower production in massive stars of $\gtrsim$ 100 M$_{\odot}$ is attributed to efficient destruction reactions.
Alpha capture by $^{23}$Na reduces sodium production, and alpha captures by $^{25}$Mg and $^{26}$Mg, seed elements of $^{27}$Al, reduce aluminum production.

In summary, a rotating model with an intermediate initial mass of 30-80 M$_{\odot}$ shows efficient production of sodium,
and a rotating model with an intermediate initial mass of 40-80 M$_{\odot}$ shows efficient production of aluminum.
Thus, these odd elements are useful to support the existence of rotationally induced mixing in the progenitor.


\section{ABUNDANCE PROFILING OF HMP STARS}

In this section, results of abundance profiling for the three most iron-deficient stars are firstly shown, then model comparisons with previous works are presented.
The best fit models are listed in Table \ref{tab_fit}, showing initial masses and rotational characteristics of models for each star.
And in Table \ref{tab_fit2}, the stellar yields of intermediate mass elements are summarized.
In the following subsections, important characteristics of observed abundance patterns and how the best fit models are selected are presented.

\subsection{SMSS 0313-6708}

\begin{figure}[htb]
		\centering
		\includegraphics[height=12cm, angle=270]{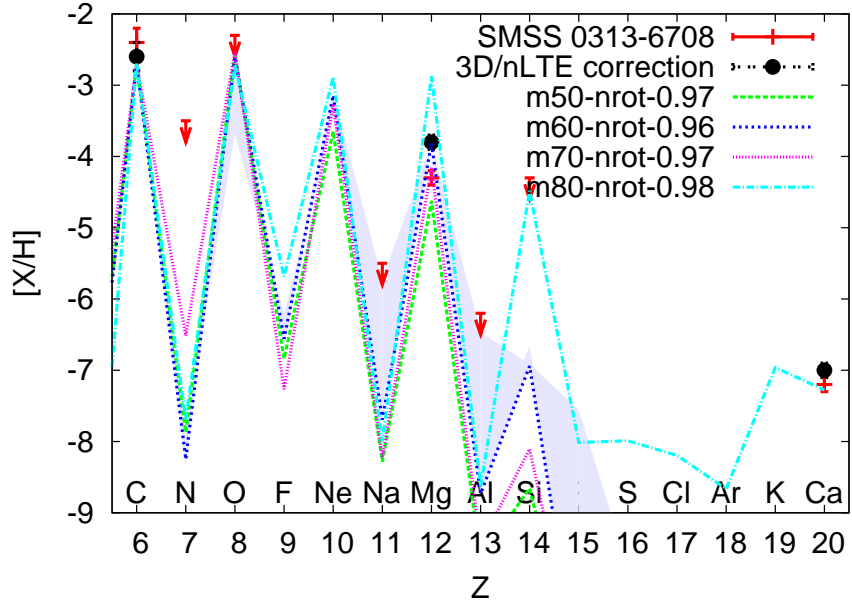}
	\caption{The abundance pattern of SMSS 0313-6708 which has [Fe/H] $<$ -7.1. Red crosses and arrows show observed values and upper limits, and black points show corrected values accounting for the effect of 3D/non-LTE stellar atmosphere, respectively. Four model yields are presented as references, they are non-rotating 50 M$_{\odot}$ with $f_{in}$=0.97 (green long-dashed), non-rotating 60 M$_{\odot}$ with $f_{in}$=0.96 (blue short-dashed), non-rotating 70 M$_{\odot}$ with $f_{in}$=0.97 (magenta dotted), and non-rotating 80 M$_{\odot}$ with $f_{in}$=0.98 (cyan dash-dotted). For the definition of $f_{in}$, see the text. A blue shadow shows the influence of changing $f_{in}$ in the range of 0.92-1.00 for the 60 M$_{\odot}$ model.}
	\label{stars-J0313}
\end{figure}

SMSS 0313-6708 is the most iron deficient star known so far.
Non-detection of iron has been reported by \citet{keller+14} and they estimate the upper limit of the iron abundance of [Fe/H] $<$ -7.1 (Thus, strictly speaking, the star is not a HMP star.).
Figure \ref{stars-J0313} shows the abundance pattern of observed and fitted data for the star in terms of [X/H].
The observed abundances and corresponding 3D and non-LTE corrections are taken from \citet{keller+14}.
Selected yield models are 50 M$_{\odot}$ with $f_{in}$=0.97, 60 M$_{\odot}$ with $f_{in}$=0.96, 70 M$_{\odot}$ with $f_{in}$=0.97, and 80 M$_{\odot}$ with $f_{in}$=0.98.
Here we define $f_{in}$ $\equiv$ $M_{in}/M_\mathrm{CO}$ as the indicator of the depth of the mass ejection.
Each selected model is non-rotating.
Labels of yield models in the figure show their parameters.
For example, m50-nrot-0.97 means the yield of a non-rotating 50 M$_{\odot}$ model with $f_{in}$=0.97.
The uncertainty range with different $f_{in}$ for the 60 M$_{\odot}$ model is shown as a blue shadow in the figure.
The range is chosen so that the observed upper limit of sodium and aluminum are reproduced, and the values are $f_{in}$=0.92-1.00.

\begin{figure}[htb]
		\centering
		\includegraphics[height=12cm, angle=270]{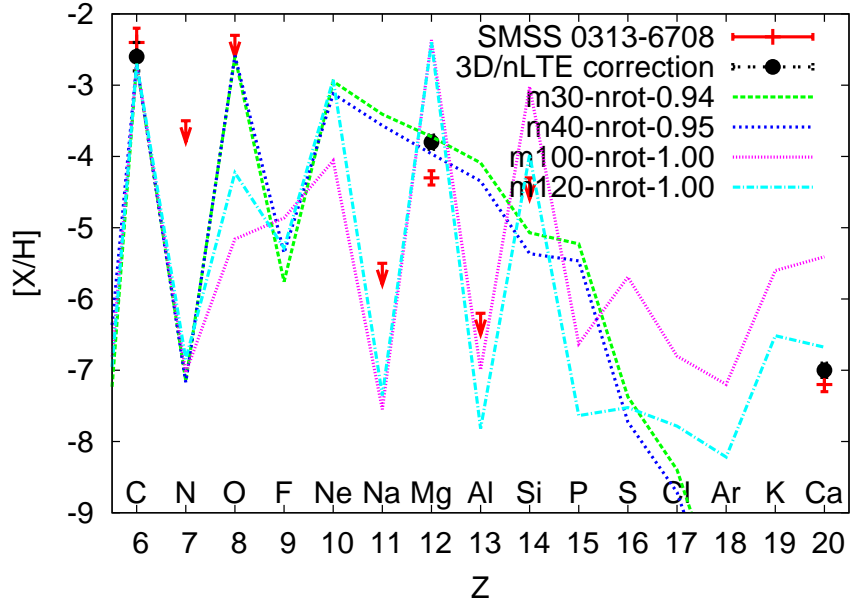}
	\caption{The abundance pattern of SMSS 0313-6708 with non-rotating model yields. For less massive 30 and 40 M$_{\odot}$ models, $f_{in}$ are selected so that the observed magnesium abundance is reproduced. For massive 100 and 120 M$_{\odot}$ models, $f_{in}$ = 1.00 is arbitrarily taken because these models cannot account for the magnesium abundance with any $f_{in}$.}
	\label{stars-J0313-o0non}
\end{figure}

\begin{figure}[htb]
		\centering
		\includegraphics[height=12cm, angle=270]{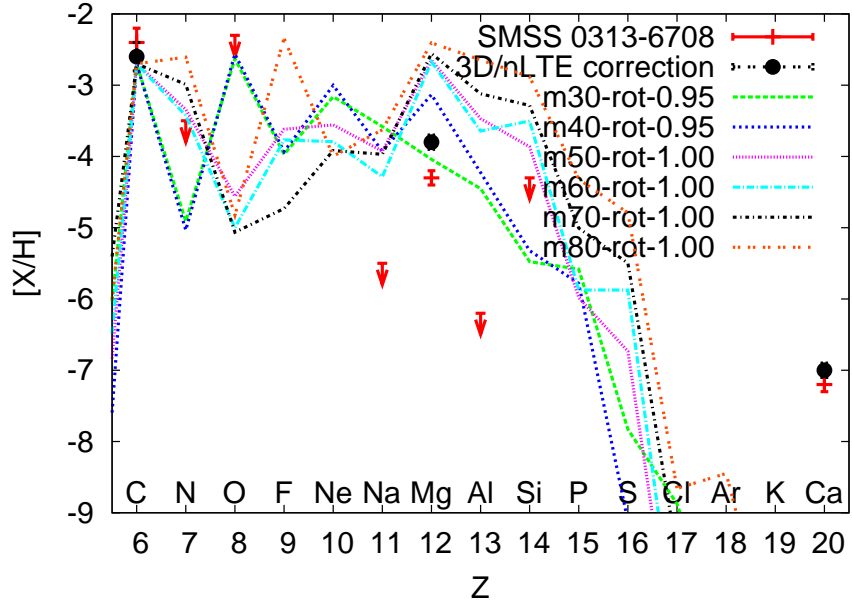}
	\caption{The abundance pattern of SMSS 0313-6708 with rotating model yields. $f_{in}$ are arbitrarily taken.}
	\label{stars-J0313-o2non}
\end{figure}

The star has small magnesium abundance compared with carbon: [Mg/C] $\sim$ -2 for the 1D-LTE value and $\sim$ -1 for the corrected value.
Heavy massive models of $\ge$ 100 M$_{\odot}$ produce much magnesium in a helium layer, and do not reproduce the observation (Fig. \ref{stars-J0313-o0non}, 100 and 120 M$_{\odot}$ models).
On the other hand, the small production ratio of magnesium can be explained by low (12-40 M$_{\odot}$) and intermediate mass (50-80 M$_{\odot}$) models.
For less massive stars, the necessary amount of magnesium can be produced by inner carbon burning layer, while magnesium production in a helium layer can account for the ratio in the case of intermediate mass stars.
Though sodium and aluminum have not yet been detected in the star, the upper limits of [Na/Mg] $<$ -1.2 and [Al/Mg] $<$ -1.9 are valuable for the abundance profiling.
The limits reject less massive models, since carbon burning produces sodium and aluminum besides magnesium (Fig. \ref{stars-J0313-o0non}, for 30 and 40 M$_{\odot}$ models).
Moreover, rotating models also produce sodium and aluminum in a helium layer (Fig. \ref{stars-J0313-o2non}), and no rotating models match with the observation.
Therefore, only non-rotating intermediate massive stars of 50-80 M$_{\odot}$ can fit the observation.
The dilution factor of the models is calculated so as to fit the observed carbon abundance, [C/H] = -2.6.
The ejected masses of carbon are 0.419-0.136 M$_{\odot}$ for the 60 M$_{\odot}$ model for $f_{in}$ = 0.92-1.00.
With the solar value of $X(\mathrm{C})$$_{\odot}$ = $2.38\times 10^{-3}$ and $X(\mathrm{H})$$_{\odot}$ = 0.7381 \citep{asplund+09} and the primordial value of  $X(\mathrm{H})_\mathrm{ISM}$ = 0.7599 \citep{steigman07}, the corresponding dilution factors become $1.78\times 10^3$-$6.09\times 10^2$.

In addition to carbon and magnesium, calcium is detected in the star as well, with the value of [Ca/H] = -7.
Since the value is very small and close to the upper limit of the iron abundance, it could be explained by another mechanism that accounts for heavy elements observed in other HMP stars.
On the other hand, the calcium production should be consistently explained with other intermediate mass elements, in case future observations reveal the overabundance of calcium to heavier elements such as iron-peak elements.
As for our models, massive stars of $>$ 80 M$_{\odot}$ produce calcium due to break-out reactions from CNO cycle at the base of hydrogen layers.
Calcium production by 80 M$_{\odot}$ model is compatible with the observation, therefore the 80 M$_{\odot}$ model can consistently explain the abundance pattern from carbon to calcium.
Though magnesium is overproduced in the helium layer, the excess may be in an uncertainty of calculations and observations.
The range of $f_{in}$ is limited by the observation, and the values are 0.94-1.02.
The 80 M$_{\odot}$ model produces $4.80\times 10^{-1}$ M$_{\odot}$ to $5.36\times10^{-2}$ M$_{\odot}$ of carbon, and corresponding dilution factors are $1.62\times 10^{3}$ to $1.91\times 10^2$.

\subsection{HE 0107-5240}

\begin{figure}[htb]
		\centering
		\includegraphics[height=12cm, angle=270]{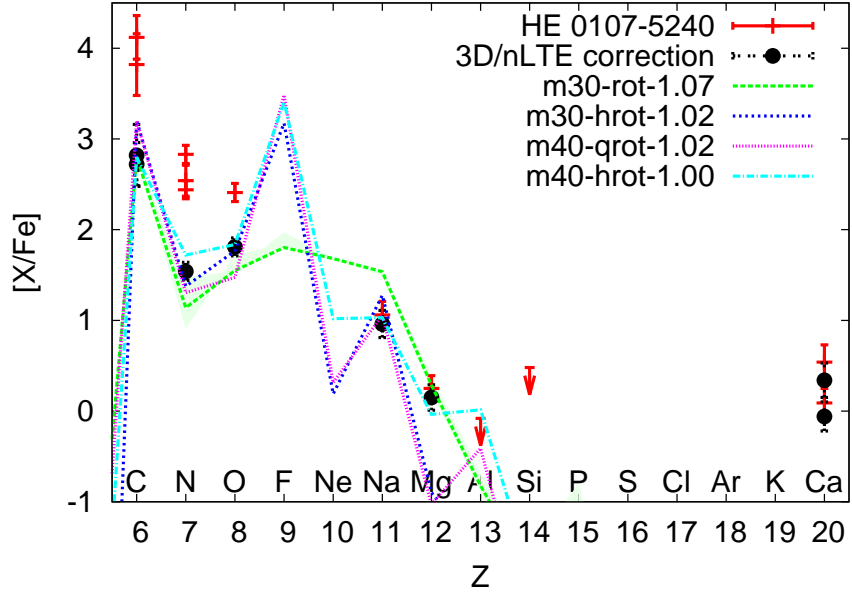}
	\caption{Same as Fig. \ref{stars-J0313}, but for HE 0107-5240. The star has [Fe/H] = -5.3. Selected model yields are rotating 30 M$_{\odot}$ with $f_{in}$=1.07 (green long-dashed), half-rotating 30 M$_{\odot}$ with $f_{in}$=1.02 (blue short-dashed), quarter-rotating 40 M$_{\odot}$ with $f_{in}$=1.02 (magenta dotted), and half-rotating 40 M$_{\odot}$ with $f_{in}$=1.00 (cyan dash-dotted). A green shadow corresponds to different $f_{in}$ models from 1.00-1.13 for the rotating 30 M$_{\odot}$ model.}
	\label{stars-HE0107-5240}
\end{figure}

HE 0107-5240 has the metallicity of [Fe/H] = -5.3, firstly reported by \citet{christlieb+02}.
Figure \ref{stars-HE0107-5240} shows the abundance pattern of the star and selected model yields.
Plotted observation points are taken from \citet{christlieb+04},  \citet{bessell+04}, \citet{bessell&christlieb05}, and from \citet{collet+06} for 3D correction.
The rotating 30 M$_{\odot}$ model is selected for the best fit model among the basic set of calculations, having the value of $f_{in}$=1.07.
In addition to the 30 M$_{\odot}$ model, we calculate slowly rotating models of 20, 30, 40 M$_{\odot}$, and find that three models of them match better with the observation.
They are a 30 M$_{\odot}$ model with a half speed of rotation, a 40 M$_{\odot}$ model with a quarter speed of rotation, and a 40 M$_{\odot}$ model with a half speed of rotation.
The fitting results are also shown in Fig. \ref{stars-HE0107-5240}.

The most important abundance ratio for the star is the very small [O/C] = -1.4.
Such a small [O/C] can only be explained by a model with a larger $f_{in}$ than unity.
This is because the small oxygen production is severely affected by mass ejection of CO core materials.
To be consistent with the small oxygen production, other abundant elements should be explained by nucleosynthesis in a helium layer in the progenitor.
The sodium abundance of [Na/C] $\sim$ -2 excludes non-rotating progenitors, because non-rotating models do not produce sodium at the helium layer.
Then the production ratio of magnesium can be used to constrain the initial mass.
Among the basic models, only the rotating 30 M$_{\odot}$ model can match the observed [Mg/C] $\sim$ -3 with sufficient production of sodium.
Less massive models lack the necessary production of magnesium, while more massive stars overproduce.
For the best fit model of rotating 30 M$_{\odot}$ case, an acceptable range of $f_{in}$=1.01-1.13 (Fig. \ref{stars-HE0107-5240}, magenta shadows) are wide.
This is because the abundance pattern from carbon to silicon is almost the same through the convective helium layer in the progenitor.
In order to fit the carbon abundance of [C/H] = -2.7, and considering the carbon yield of $7.20 \times 10^{-2}$-$1.90 \times 10^{-2}$ M$_{\odot}$ by the rotating 30 M$_{\odot}$ model with $f_{in}$ = 1.01-1.13, the dilution factor becomes $7.84\times 10^2$-$2.23\times 10^2$.

Since the sodium production of the rotating 30 M$_{\odot}$ model slightly exceeds the observed value, we calculate slowly rotating models in addition.
Owing to the slower rotation, these models produce less sodium than the basic rotating model, and thus match with the observation.
The 40 M$_{\odot}$ model with a half speed of rotation is the best fit model.
The small abundance of [Mg/Fe] $\sim$ 0 may be explained by other pollution mechanisms discussed later.
Therefore, the half-rotating 30 M$_{\odot}$ model and the quarter-rotating 40 M$_{\odot}$ model may be also compatible with the observation.

\subsection{HE 1327-2326}

\begin{figure}[htb]
		\centering
		\includegraphics[height=12cm, angle=270]{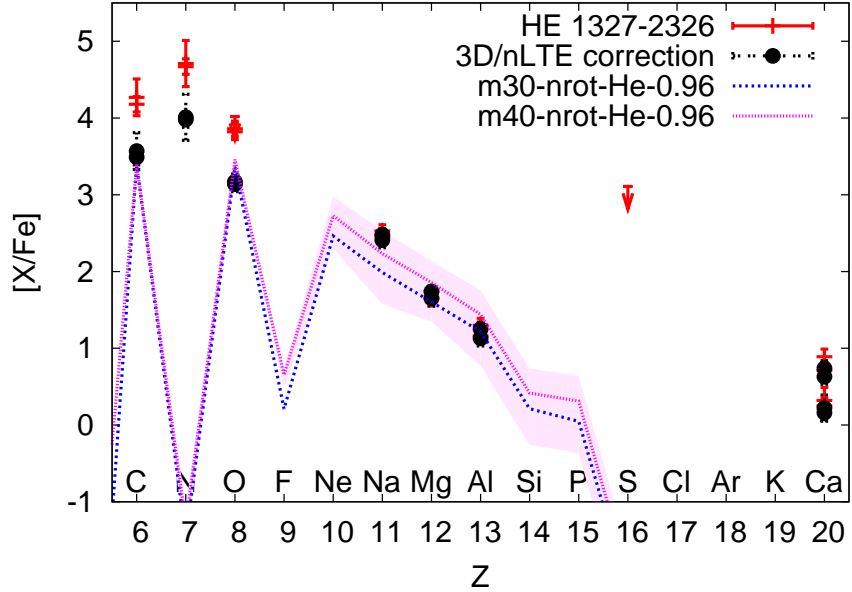}
	\caption{Same as Fig. \ref{stars-J0313}, but for HE 1327-2326. The star has [Fe/H] = -5.7. Model yields of non-rotating 30 M$_{\odot}$ with $f_{in}$=0.96 (blue short-dashed) and non-rotating 40 M$_{\odot}$ with $f_{in}$=0.96 (magenta dotted) are presented. A magenta shadow corresponds to different $f_{in}$ models from 0.95-0.97 for the 40 M$_{\odot}$ model.}
	\label{stars-HE1327-2326-o0}
\end{figure}

\begin{figure}[htb]
		\centering
		\includegraphics[height=12cm, angle=270]{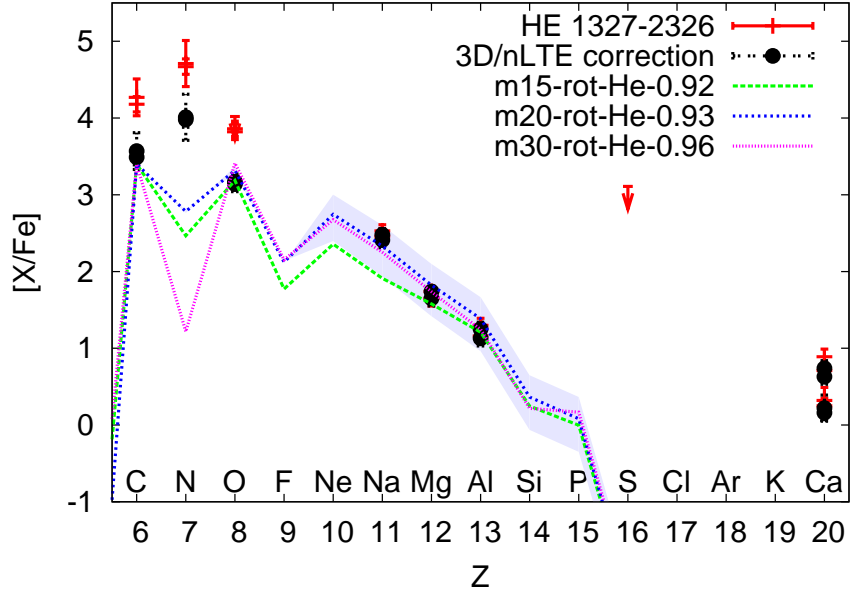}
	\caption{Same as Fig. \ref{stars-J0313}, but for HE 1327-2326. Three model yields of rotating 15 M$_{\odot}$ with $f_{in}$=0.92 (green long-dashed), rotating 20 M$_{\odot}$ with $f_{in}$=0.93 (blue short-dashed), and rotating 30 M$_{\odot}$ with $f_{in}$=0.96 (magenta dotted) are presented. A blue shadow corresponds to different $f_{in}$ models from 0.92-0.94 for the 20 M$_{\odot}$ model.}
	\label{stars-HE1327-2326-o2}
\end{figure}

HE 1327-2326 is an HMP star with [Fe/H] = -5.7, reported by \citet{frebel+05}.
Figures \ref{stars-HE1327-2326-o0} and \ref{stars-HE1327-2326-o2} are the same as Fig. \ref{stars-J0313}, but for HE 1327-2326.
These figures have the same observation points, but have different model fitting.
In the former figure, non-rotating models are presented, while the latter shows rotating models.
Plotted observation points are taken from \citet{aoki+06}, \citet{frebel+06, frebel+08}, and \citet{bonifacio+12}.
For 3D correction, results in \citet{collet+06} are applied onto data obtained by \citet{aoki+06} and \citet{frebel+06}.

The abundance ratios of [O/C] and [Mg/C] are used to constrain progenitor's initial masses.
Since the star shows negative [O/C], the ejection of the inner matter of the carbon oxygen core should be limited.
Then, the magnesium abundance of [Mg/C] $\sim$ -2.7 can be used to constrain the initial mass.
Massive stars of $\ge$ 50 M$_{\odot}$ overproduce magnesium in outer helium layers.
On the other hand, low mass models of $\leq$ 40 M$_{\odot}$ can explain the abundance, ejecting the outer edge of the convective carbon burning region.
In this case, because sodium, magnesium, and aluminum are produced in the same region, observed ratios of [(Na, Mg, Al)/C] can be simultaneously explained.
However, the smallest 12 M$_{\odot}$ models produce less sodium and fail to explain the sodium ratio, non-rotating 15, 20 M$_{\odot}$ models suffer from proton ingestion and overproduce calcium.
Therefore, intermediate mass stars of 30-40 M$_{\odot}$ for non-rotating models and 15-30 M$_{\odot}$ for rotating models give consistent yield to the observation.
The best fit models are 40 M$_{\odot}$ for non-rotating and 20 M$_{\odot}$ for rotating models.
Because $M_{in}$ in these models are set to the edges of carbon convective regions, acceptable widths in terms of $f_{in}$ are very narrow.
They are shown in Figs. \ref{stars-HE1327-2326-o0} and \ref{stars-HE1327-2326-o2} as colored shadows, corresponding to $f_{in}$= 0.95-0.97 for the non-rotating 40 M$_{\odot}$ model and to $f_{in}$ = 0.92-0.94 for the rotating 20 M$_{\odot}$ model.
The carbon yields are 0.198-0.170 M$_{\odot}$ for the non-rotating 40 M$_{\odot}$ model with $f_{in}$ = 0.95-0.97, and 0.175-0.161 M$_{\odot}$ for the rotating 20 M$_{\odot}$ model with $f_{in}$ = 0.92-0.94.
The star has a carbon abundance of [C/H] = -2.2, and corresponding dilution factors of the two models become $5.00\times 10^2$-$4.32\times 10^2$ for the non-rotating 40 M$_{\odot}$ model and $7.92\times 10^2$-$7.35\times 10^2$ for the rotating 20 M$_{\odot}$ model.

HE 1327-2326 shows significant enhancements of nitrogen.
Although none of our calculation presented in this work do not consistently match with the nitrogen abundance, a weak supernova from a rotating progenitor may be able to explain the observation.
Our rotating models only include moderate rotators of $v_{rot}/v_k \sim 0.15$, and the best model of 20 M$_{\odot}$ yields about 1/10 of observed nitrogen.
Therefore, one possibility to account for the large production is fast rotation in which highly effective internal mixing will take place.
Also, a rotating star with a very small but non-zero metallicity is known to have a large enhancement in nitrogen production.
Comparing the results by \citet{ekstroem+08} and \citet{hirschi07}, models with a metallicity of $10^{-8}$ show larger nitrogen production than Pop III models.
\citet{ekstroem+08} has explained this trend as a consequence of existence and absence of CNO elements at its birth.
This is because, a metal poor progenitor with CNO elements can support the structure by the CNO cycle from the first ignition of hydrogen.
At the end of the core hydrogen burning phase, this results in faster core rotation and thus more effective internal mixing.

An isotopic ratio of $^{12}$C/$^{13}$C is useful to distinguish models.
Theoretically, nitrogen in a low mass metal poor star can be synthesized by an internal process called the helium-flash driven deep mixing (He-FDDM).
In this scenario, a convective region powered by a shell helium-flash penetrates into the hydrogen envelope, resulting in nucleosynthesis of nitrogen \citep[e.g.,][]{fujimoto+90, suda+04}.
However, the observed isotopic ratio of $^{12}$C/$^{13}$C $>$ 5 disfavors the scenario in the case of HE 1327-2326, since the theory predicts the equilibrium value of the CN cycle, $^{12}$C/$^{13}$C $\sim$ 3-4 \citep{picardi+04, weiss+04, aoki+06}.
In our model, the isotopic ratio is 21.5 for the rotating 20 M$_{\odot}$ progenitor, and agrees with the observation.
However, the value will be reduced by a factor of 1/10 when a required amount of $^{14}$N is produced, since both of the elements are simultaneously produced by the CNO cycle.

\subsection{Model Comparison}

\subsubsection{SMSS 0313-6708}

\paragraph{Model Comparison with \citet{keller+14}}

The model described in \citet{keller+14} is similar to ours.
They attribute the observed abundances to a 1.8 $\times$ $10^{51}$ erg supernova explosion from a 60 M$_{\odot}$ Pop III progenitor.
The explosion energy is small compared to the relatively large mass of the progenitor, and the initial mass is in the range of our results of 50-80 M$_{\odot}$.

There is a small difference on the origin of calcium between the model by \citet{keller+14} and ours.
\citet{keller+14} has reported that the calcium is produced by the break-out reactions from the CNO cycle during the stable hydrogen burning phase.
On the other hand, the central temperature during the main sequence stage never reaches the allowed value for the break-out reactions in our models.
Instead, calcium is produced at the base of the hydrogen burning shell in massive non-rotating models of $\ge$ 80 M$_{\odot}$ after the carbon burning stage.
In addition, models of $\ge$70 M$_{\odot}$ in \citet{keller+14} are rejected, since these models do not reproduce the carbon enhancement and overproduce nitrogen.
On the other hand, the reason for the rejection of the heavier models of $\ge$100 M$_{\odot}$ in our models is due to the over abundance of magnesium.

\paragraph{Model Comparison with \citet{ishigaki+14}}

Among the models reported in \citet{ishigaki+14}, a $10^{52}$ erg explosion from a 25 M$_{\odot}$ progenitor consistently reproduces the observed abundance patterns of [(C, Na, Mg, Al)/Ca].
Considering the extremely low escape fraction of $10^{-5.8}$ of the mixing-fallback region, the model will have a similar structure of matter ejection to our model.
However, assumed explosion energies are different.

In the 25 M$_{\odot}$ model, the explosion energy is so large that the explosive helium burning takes place at the base of the helium layer.
Owing to the magnesium production by the explosive nucleosynthesis, observed magnesium abundance can be explained by the model.
On the other hand, no explosive nucleosynthesis is assumed to occur in the weak explosions, and magnesium in our models is produced during the pre-collapse stages.
The difference in the explosion energy will affect the resulting metal pollution.
We later discuss different efficiencies of the metal pollution from different explosion energies in Section 5.4.

To distinguish these models with different explosion energies, the oxygen abundance may be useful.
Our best fit models suggest that the oxygen abundance of the star may be [O/C] $\sim$ 0, while the 25 M$_{\odot}$, $10^{52}$ erg explosion model by \citet{ishigaki+14} have much lower value of $\sim$ -2.
These values are due to different inner boundary masses of ejection.
The predicted value of [O/C] $\sim$ 0 by ours is just under the observed upper limit, thus the detection of oxygen by future observations may be useful to constrain the supernova model.

\subsubsection{HE 0107-5240}

\paragraph{Model Comparison with \citet{iwamoto+05}}

As discussed earlier, a large mass coordinate of the fallback boundary is necessary to explain the small [O/C] ratio by a supernova ejection.
Accordingly, the model presented by \citet{iwamoto+05} has a large mass cut (that is equivalent to our $M_{in}$) of 6.3 M$_{\odot}$ for a 25 M$_{\odot}$ progenitor.
Thus the large inner boundary and the initial mass are compatible with ours.

The difference between the two models is origins of sodium and magnesium.
They attribute production of these elements to very small ejection of processed inner matter in the CO core with a small escape fraction of 1.2 $\times$ $10^{-4}$.
A high degree of fine-tuning of the escape fraction and of the mass cut will be needed in the model.
On the other hand, our model produces those intermediate mass elements by rotationally induced nuclear reactions, which take place in a wide range of initial parameters of the initial mass and the initial rotational velocity.
The rotating models reproduce similar abundance yield under a wide parameter range in $M_{in}$, since the abundance distributions are almost constant within a convective helium layer.
Therefore, the rotating model will be more robust to explain the observation than the models with inefficient inner matter ejections.

\paragraph{Model Comparison with \citet{limongi+03}}

In \citet{limongi+03}, two supernovae are considered to explain the metal pollution, i.e., the observed abundance is explained as a superposition of the yields.
The first one is a less energetic 35 M$_{\odot}$ supernova with a mass cut of 9.4 M$_{\odot}$ contributing elements from carbon to magnesium, and the other one is a typical 15 M$_{\odot}$ supernova with an iron production of 5.6$\times$ 10$^{-2}$ M$_{\odot}$ contributing heavier elements.

The strategy dividing elements into lighter and heavier elements is similar to ours, and the progenitor mass and the considered explosion for explaining the lighter elements are also compatible with our models.
The largest difference arises in a production mechanism of sodium.
In the 35 M$_{\odot}$ model in \citet{limongi+03}, sodium production is attributed to proton ingestion into a helium burning shell.
Of course this process is one possibility, but the characteristics of proton ingestion are too complicated to be properly treated as we discuss above.
On the other hand, sodium production by rotationally induced reactions may be more robust, since it needs just an efficient internal mixing due to stellar rotation.


\section{DISCUSSION}

\subsection{Mechanisms for a Weak Explosion}

We assume that the supernova explosion is {\it weak} in its energy to explain the abundance of HMP stars.
And the mechanisms that can account for the weak explosion are not unique.
Until now, two different models are considered to account for the weak explosion; a spherical explosion and a jet-like explosion \citep{tominaga+07b}.
In addition to the two models, {\it failed supernovae}, which eject their outer layer due to the reduction of gravitational mass by high energy neutrino emission \citep{nadezhin80, lovegrove&woosley13}, may also be compatible with the weak explosion.

Even for a simple one dimensional explosion, however, it is difficult to relate $M_{in}$ with explosion energies.
This is because $M_{in}$ is very sensitive to the treatment of the engine of the explosion.
With different positions and models of energy ingestion (e.g. the piston model and the thermal bomb model), different $M_{in}$ are resulted for the same low explosion energy.
Moreover, multi dimensional calculations may result in quantitatively different matter ejection.
In a two dimensional calculation for the jet-like explosion, inner material placed in an off-axis region accretes onto the formed compact object \citep{tominaga+07a, tominaga09}.
Since the accretion depends on several jet parameters, such as the jet opening angle, the energy injection rate, and the total injected energy, degeneracies may arise in these parameters.
As for the failed supernova, the explosion mechanism is totally different from the above two models.
Because of the loss of the proto-neutron star's gravitational mass by neutron emission, a weak shock is launched by acceleration in the core and propagates outward.
Successful matter ejections are reported for 15 and 25 M$_{\odot}$ red giants by \citet{lovegrove&woosley13}.
The mechanism may be only applicable to low mass red giants, because a small core mass in the progenitor is important both for the long duration of gravitational mass reduction and for the shock propagation through the core.
What realistic models can account for the explosion with a large accretion?
Further investigation on this topic is needed.

\subsection{Origins of Heavy Elements}

In this work, we do not specify the origins of heavy elements that exist in HMP stars.
These elements will be synthesized by an explosive nucleosynthesis, and do not give severe constraints on the progenitor's initial parameters \citep{umeda&nomoto05}.
However, to consistently explain all of the abundance observations, the origin of the heavy elements should be considered.
Here we discuss three possible scenarios to account for the metal pollution.

The first possibility is an ejection of a tiny fraction of the inner processed materials by the same explosion.
This is a usual assumption made in the mixing-fallback model.
Since the observed abundance of heavy elements is very small, the escape fraction of the inner matter becomes very small as well.
In order to represent the small escape fraction, a realistic modeling may require fine-tunings of explosion parameters.
The second one is metal pollution by another supernova explosion \citep{limongi+03}.
Elements with abundances of [X/Fe] $\sim$ 0 are expected to be explained naturally by this scenario, since the usual supernova will show [X/Fe] $\sim$ 0 abundances.
Compared with single explosion models, the double explosion model has a lot of parameters to specify the model, such as mass ratio between the two progenitors, explosion energies, the time delay between the two explosions, and efficiencies of metal pollution by each explosion.
The last scenario we discuss here is ISM accretion onto the formed second generation star \citep{yoshii81}.
Only a tiny amount of metals is required to be accreted during the long lifetime of the star.
However, still both a realistic theory of the accretion and an observational support of the phenomena are lacking \citep[however, see][]{hattori+14}.

\subsection{Proton Ingestion}

In non-rotating 15 and 20 M$_{\odot}$ models in our calculation, convection in a helium layer penetrates the boundary between helium and hydrogen layers.
As a consequence, fresh fuel of hydrogen is mixed into the high temperature region in the helium layer.
Resulting energy generation powers the convection to grow and the entire former helium layer and hydrogen envelope are covered by a single convective region.
The CNO cycle accounts for the energy generation, thus, nitrogen is synthesized and distributed in the convective region at the same time.

Such a phenomenon is often reported in the literature \citep[e.g.,][]{ekstroem+08, heger&woosley10, limongi&chieffi12, yoon+12}, in which stellar evolution of Pop III stars is calculated.
Indeed, the hydrogen ingestion significantly affects both the chemical distribution and the envelope structure.
Some of them are reported to produce enough nitrogen or sodium to explain observed abundances \citep{iwamoto+05, limongi+03}.
The resulting inflation of the envelope may affect dynamics during the last explosion \citep{heger&woosley10}.

However, it is highly uncertain whether the convective penetration over the boundary and/or such a powerful mixing, by which energetic nuclear fuel are transported to deep inside the star, could occur in reality or not.
In a one-dimensional treatment, the relatively small entropy barrier at the boundary of helium and hydrogen layers may support the occurrence of the ingestion \citep{fujimoto+90}, but there still exists a $\mu$ barrier representing a composition jump that contrarily inhibits the mixing.
Moreover, in order to accurately treat the dynamical behavior of convective boundaries, multi dimensional modelings are needed \citep[e.g.,][]{meakin&arnett07}.
For canonical one dimensional calculations, not all of massive Pop III models show the ingestion, and the effects after the ingestion are not similar among the simulations.
Apparently the characteristics of the phenomenon severely depend on the numerical settings.
Proton ingestion may have many important consequences for Pop III stellar evolution, however, more sophisticated treatments than simple one dimensional calculations are needed to reveal the nature.

\subsection{Explosion Energy Dependence of Metal Pollution in the Primordial Gas Clouds}

Explosion energy is important to determine the absolute value of metal abundance in the second generation stars.
According to one dimensional calculations by \citet{kitayama&yoshida05}, since irradiation by the central star expands the ambient primordial gas in advance of the explosion, even a weak explosion of $10^{50}$ erg blows away all gas in the halo of $10^5$-$10^6$ M$_{\odot}$.
In a recent three dimensional calculation by \citet{ritter+12}, about a half of the supernova ejecta of $10^{51}$ erg escapes from the host halo of $\sim$$10^{6}$ M$_{\odot}$, while the rest is trapped by the high density flow region.
Formation of second generation stars requires high density metal polluted gas.
An escaping mass will sweep up halo gas and will become dilute inter halo gas, having too low metal abundance to form second generation stars.
Hence, the first possibility for the formation site is a metal polluted gas cloud that survives the Pop III supernova explosion.
It will collapse again due to gravity of the dark matter host halo.

Adopting the result of \citet{ritter+12}, the resulting carbon abundance in the surviving gas can be deduced from an order-of-magnitude estimate.
The escape fraction of the supernova ejecta is defined as $f_{esc}$.
In the case of $10^{51}$ erg explosion, the value is about 0.5, and the fraction will be an increasing function with the explosion energy.
A total amount of ejected carbon is set as $M_\mathrm{C}$.
The typical value depends on what kind of explosion is assumed, and it is 0.1 M$_{\odot}$ for the weak supernova model.
The ambient gas mass is about $\sim$20\% of the dark matter mass, and the value is $\sim$2$\times$$10^{5}$ M$_{\odot}$ for cosmologically typical star forming halos of $10^{6}$ M$_{\odot}$ \citep{bromm13}.
Not all of the gas may be polluted by the explosion, and the fraction of polluted gas is defined as $f_{pol}$.
In \citet{ritter+12}, the total mass of polluted gas will be $M_{pol}\sim4\times10^4$ M$_{\odot}$, and thus $f_{pol}$ becomes 0.2.
Hence, the resulting carbon abundance can be written as
\begin{eqnarray*}
X(\mathrm{C}) &=& \frac{(1-f_{esc}) \times M_\mathrm{C} }{f_{pol} \times 2\cdot10^5}\\
&=& 1.25 \times 10^{-6} \Bigl( \frac{M_\mathrm{C}}{0.1} \Bigl) \Bigl( \frac{0.2}{f_{pol}} \Bigl) \Bigl( \frac{1 - f_{esc}}{0.5} \Bigl).
\end{eqnarray*}
This is equivalent to [C/H] = $ -3.29 + \mathrm{log}[ (M_\mathrm{C}/0.1) (0.2/f_{pol}) (1 - f_{esc})/0.5 ]$ when the solar values of $X(\mathrm{H})_\odot$ = 0.7381 and $X(\mathrm{C})_\odot$ = $2.38\times10^{-3}$, and the primordial value of $X(\mathrm{H})_\mathrm{ISM}$ = 0.7599 are adopted and hydrogen yield by the supernova explosion are neglected.

For weakly energetic explosions, [C/H] will increase due to a decrease in $f_{pol}$.
On the other hand, an increase in $f_{esc}$ will reduce [C/H] for highly energetic explosions.
Since the expected value of $10^{51}$ erg is lower than the observed value of [C/H] $\sim$ -2.6 for SMSS 0313-6708, weakly energetic supernovae may be the most likely candidate for the progenitor.
Of course, large uncertainty exists in this estimate, for example, a three dimensional dense flow structure may help the ejecta to survive a highly energetic explosion as reported in the $10^{51}$ erg supernova.


\section{Summary and Conclusion}

The main purpose of this work is to obtain new knowledge of abundance yields of Pop III supernovae that can be used to constrain the characteristics of Pop III stars.
We calculate Pop III progenitor evolution in a wide range of initial parameters, and calculate the stellar yields with the assumption of a weak explosion.
The initial mass range is from 12 to 140 M$_{\odot}$ so that the whole mass region for core collapse supernovae is covered.
Stellar rotation is newly included in the progenitor calculation, resulting in diverse nucleosynthesis due to efficient internal matter mixing.

We show that various abundance distributions arise in outer regions in calculated models.
Massive models of $\ge$ 40 M$_{\odot}$ for rotating and $\ge$ 60-80 M$_{\odot}$ for non-rotating cases show both magnesium and silicon enhancement in their helium layers.
These enhancements are due to efficient alpha capture reactions in the region.
As for rotating models, owing to rotationally induced mixing, abundant nitrogen is produced in the hydrogen burning shell at first.
Alpha capture reactions onto nitrogen take place in later evolutionary phases, resulting in neutron emission and nucleosynthesis of sodium and aluminum.
For non-rotating heavy massive stars of $\ge$ 80 M$_{\odot}$, calcium production occurs in the hydrogen burning shell owing to break-out of the CNO cycle.

We show results of abundance profiling to the three most iron deficient stars.
The abundance pattern of SMSS 0313-6708 can be explained by non-rotating massive 50-80 M$_{\odot}$ models with large inner boundaries of ejections, $f_{in}$ $\sim$ 0.92-1.00.
The non-rotating 60 M$_{\odot}$ model gives the best explanation to the observed low [Mg/C] with upper limits on [(Na, Al)/C], while the small abundance of [Ca/C] can be consistently explained by the 80 M$_{\odot}$ model.
HE 1327-2326 has a small [O/C] and an interesting abundance sequence of [(Na, Mg, Al)/C].
These abundances are consistently explained by both rotating and non-rotating 15-40 M$_{\odot}$ models with ejections from the outer edge of the carbon convection regions, $f_{in}$ $\sim$ 0.92-0.97.
To explain the large abundance of [N/C], other origins than the single explosion may be needed.
Small abundances of [(N, O, Na)/C] in HE 0107-5240 can be consistently explained by a rotating 30 M$_{\odot}$ model, with a wide acceptable range of $f_{in}$ $\sim$ 1.01-1.13.
Additionally calculated 30 and 40 M$_{\odot}$ models with slow rotation show much better fitting for the sodium abundance.

Finally, we compare our results with other theoretical fittings in the literature.
For SMSS 0313-6708, models in \citet{keller+14} and in \citet{ishigaki+14} are compared with ours.
The model by \citet{keller+14} is similar to our models, while they have different origins of calcium.
The most important difference between ours and the model by \citet{ishigaki+14} is the assumed explosion energy, and our weakly energetic explosion may be more plausible to explain the observed carbon abundance.
For HE 0107-5240, models in \citet{iwamoto+05} and in \citet{limongi+03} are compared with ours.
The model by \citet{iwamoto+05} does not yield sodium at the outer region, thus they assume highly inefficient inner matter ejection to account for the sodium production.
The 35 M$_{\odot}$ model in \citet{limongi+03} does yield sodium at its outer layer, however, they attribute the sodium production to the proton ingestion.
In our models, rotationally induced mixing naturally results in sodium production, and thus the production mechanism may be more robust than others.

In conclusion, we constrain the initial parameters of the Pop III mother stars for the three most iron-deficient stars.
Not only the deficiency of iron, but also the enhancement of intermediate mass elements of carbon, nitrogen, oxygen, sodium, and magnesium is an important feature of these stars.
We found that this peculiar abundance feature is also useful to constrain the initial parameters of the progenitor star.
The small abundances of [O/C] are well explained by weak supernova models, the progenitor masses can be constrained by [(Mg, Si)/C], and the [(Na, Al)/C] are used to constrain the progenitor rotation.
Similar analysis of the abundance profiling will be applicable to other carbon-enhanced HMP stars, which will be discovered by future observations.
The results may constrain the characteristics of the primordial stars in the early universe.

The authors are grateful to the referee for many invaluable comments, which help to improve the paper.
The authors would like to thank N. Tominaga and S. Hirano for many fruitful discussions and N. Iwamoto for giving us critical advices.
We thank Aaron C. Bell for the careful reading of the manuscript.
The author K. T. is supported by Research Fellowships of Japan Society for the Promotion of Science (JSPS) for Young Scientists.
This work has been supported in part by JSPS KAKENHI grant Nos. 22540246 and 26400271.


\begin{table}[htdp]
	\begin{center}
	\begin{tabular}{cc|cc}
\hline
\hline
	Element & $A$ & Element & $A$ \\
\hline
	n	&	1	&	Ar	&	33-42	\\
	H	&	1-3	&	K	&	36-43	\\
	He	&	3-4	&	Ca	&	37-48	\\
	Li	&	6-7	&	Sc	&	40-49	\\
	Be	&	7-9	&	Ti	&	41-51	\\
	B	&	8-11	&	V	&	44-52	\\
	C	&	11-14	&	Cr	&	46-55	\\
	N	&	12-15	&	Mn	&	48-56	\\
	O	&	13-20	&	Fe	&	50-61	\\
	F	&	17-21	&	Co	&	54-62	\\
	Ne	&	18-24	&	Ni	&	56-66	\\
	Na	&	20-25	&	Cu	&	59-67	\\
	Mg	&	21-27	&	Zn	&	62-70	\\
	Al	&	23-29	&	Ga	&	65-73	\\
	Si	&	24-32	&	Ge	&	69-76	\\
	P	&	27-34	&	As	&	71-77	\\
	S	&	29-36	&	Se	&	73-79	\\
	Cl	&	31-38	&	Br	&	76-80	\\
\hline
	\end{tabular}
	\end{center}
	\caption{Isotopes included in the nuclear reaction network for stellar evolution.}
	\label{tab_isotope}
\end{table}

\begin{table}[htdp]
	\begin{center}
	\begin{tabular}{ccccccccccc}
\hline
\hline
	$M_{ini}$ & $M_{fin}$ & $v_{rot}$ & $v_{rot}/v_k$ & $\tau_\mathrm{H}$ & $\tau_{fin}$ & $M_\mathrm{Fe}$ & $M_\mathrm{CO}$ & $\Delta M_\mathrm{He}$ & log $T_{base, \mathrm{He}}$ & log $T_{base, \mathrm{H}}$ \\
\hline
	12	&	12	&	0	&	0	&12.46	&14.80	&	1.508	&	2.516	&	0.897	&	8.569	&	7.729	\\
	15	&	15	&	0	&	0	&	9.94	&12.62	&	1.419	&	3.596	&	0.145	&	8.151	&	7.660	\\
	20	&	20	&	0	&	0	&	7.86	&	9.40	&	1.644	&	5.730	&	0.126	&	8.232	&	7.832	\\
	30	&	30	&	0	&	0	&	5.62	&	6.52	&	1.845	&	10.28	&	0.252	&	8.461	&	8.171	\\
	40	&	40	&	0	&	0	&	4.39	&	5.02	&	2.206	&	15.07	&	1.571	&	8.564	&	8.303	\\
	50	&	50	&	0	&	0	&	3.70	&	4.26	&	2.454	&	19.35	&	1.537	&	8.639	&	8.422	\\
	60	&	60	&	0	&	0	&	3.22	&	3.75	&	2.631	&	23.63	&	2.224	&	8.746	&	8.484	\\
	70	&	70	&	0	&	0	&	3.04	&	3.46	&	2.755	&	28.95	&	1.897	&	8.706	&	8.520	\\
	80	&	80	&	0	&	0	&	2.86	&	3.25	&	3.799	&	33.81	&	2.111	&	8.811	&	8.617	\\
	100	&100	&	0	&	0	&	2.60	&	2.94	&	4.748	&	43.60	&	2.608	&	8.840	&	8.651	\\
	120	&120	&	0	&	0	&	2.44	&	2.75	&	4.353	&	53.33	&	2.976	&	8.816	&	8.631	\\
	140	&140	&	0	&	0	&	2.31	&	2.62	&	12.46	&	63.18	&	3.339	&	8.951	&	8.766	\\
\hline
	12	&	12	&	210	&	0.15	&13.05	&14.61	&	1.452	&	2.448	&	1.374	&	8.475	&	7.281	\\
	15	&	15	&	220	&	0.15	&10.64	&11.90	&	1.520	&	3.674	&	1.512	&	8.221	&	6.636	\\
	20	&	20	&	230	&	0.15	&	8.63	&	9.49	&	1.541	&	6.191	&	1.628	&	8.030	&	6.210	\\
	30	&	30	&	250	&	0.15	&	5.72	&	6.42	&	2.001	&	11.10	&	1.955	&	8.435	&	7.742	\\
	40	&39.74	&	250	&	0.15	&	5.07	&	5.87	&	2.604	&	16.45	&	2.387	&	8.776	&	8.159	\\
	50	&49.28	&	270	&	0.15	&	4.44	&	5.14	&	3.698	&	23.90	&	2.790	&	8.832	&	8.209	\\
	60	&58.84	&	270	&	0.14	&	3.92	&	4.46	&	4.084	&	28.58	&	3.388	&	8.844	&	8.240	\\
	70	&68.57	&	280	&	0.14	&	4.04	&	4.58	&	4.648	&	33.74	&	4.004	&	8.851	&	8.230	\\
	80	&77.39	&	280	&	0.14	&	3.80	&	4.36	&	6.017	&	42.48	&	4.249	&	8.872	&	8.266	\\
	100	&95.94	&	280	&	0.13	&	3.09	&	3.59	&	7.644	&	50.43	&	5.833	&	8.910	&	8.371	\\
	120	&114.89	&	280	&	0.13	&	2.88	&	3.27	&	16.62	&	59.58	&	6.151	&	8.980	&	8.478	\\
	140	&134.38	&	270	&	0.12	&	2.58	&	2.94	&	21.77	&	70.13	&	7.602	&	9.011	&	8.468	\\
\hline
	20	&	20	&	59	&	0.04	&	7.96	&	8.92	&	1.473	&	6.069	&	1.433	&	7.962	&	6.247	\\
	30	&	30	&	64	&	0.04	&	5.41	&	6.35	&	2.049	&	10.84	&	1.953	&	8.469	&	7.437	\\
	40	&	40	&	66	&	0.04	&	4.33	&	4.88	&	2.539	&	15.31	&	2.002	&	8.590	&	8.186	\\
	20	&	20	&	120	&	0.08	&	7.99	&	8.90	&	1.499	&	5.808	&	1.618	&	8.065	&	6.214	\\
	30	&	30	&	130	&	0.08	&	5.84	&	6.49	&	1.572	&	10.60	&	1.788	&	8.369	&	7.791	\\
	40	&	40	&	130	&	0.08	&	4.65	&	5.38	&	2.287	&	16.79	&	1.990	&	8.634	&	8.106	\\
\hline
	\end{tabular}
	\end{center}
	\caption{Model properties. $M_{ini}$ and $M_{fin}$ are the initial and final masses ; $v_{rot}$ and $v_k \equiv \sqrt{GM/R}$ are the surface rotation velocity and the surface Kepler velocity at the zero age main sequence; $\tau_\mathrm{H}$ and $\tau_{fin}$ are the hydrogen burning duration and the lifetime; $M_\mathrm{Fe}$, $M_\mathrm{CO}$ and $\Delta M_\mathrm{He}$ are the iron core mass, the carbon-oxygen core mass, and the helium shell mass at the end of the calculation; and $T_{base, \mathrm{He}}$ and $T_{base, \mathrm{H}}$ are the base temperatures of the helium shell and the hydrogen envelope at the end of the calculation. Masses are in M$_{\odot}$, velocities are in km/sec, time are in $10^6$ yr, and temperatures are in K.}
	\label{tab_models}
\end{table}

\clearpage

\clearpage

\begin{table}[htdp]
	\begin{center}
	\begin{tabular}{cccccc}
\hline
\hline
	Object			& [Fe/H]	& $M_{ini}$ 	& 	$f_{in}$						&	Rotation		&	Dilution Factor\\
\hline
	SMSS 0313-6708	& $<$ -7.1	&	50-80	&	0.96 $\pm$ 0.04 (60 M$_{\odot}$)	&	non-rotating	&	$1.78 \times 10^3 - 6.09 \times 10^2$\\
					& 		&			&	0.98 $\pm$ 0.04 (80 M$_{\odot}$)	&	non-rotating	&	$1.62 \times 10^3 - 1.91 \times 10^2$\\
	HE 0107-5240		&	-5.3	&	30-40	&	1.07 $\pm$ 0.06 (30 M$_{\odot}$)	&	rotating		&	$7.84 \times 10^2 - 2.23 \times 10^2$\\
	HE 1327-2326		&	-5.7	&	20-40	&	0.96 $\pm$ 0.01 (40 M$_{\odot}$)	&	non-rotating	&	$5.00 \times 10^2 - 4.32 \times 10^2$\\
					&		&	15-30	&	0.93 $\pm$ 0.01 (20 M$_{\odot}$)	&	rotating		&	$7.92 \times 10^2 - 7.35 \times 10^2$\\
\hline
	\end{tabular}
	\end{center}
	\caption{Summary of abundance profiling.}
	\label{tab_fit}
\end{table}

\begin{table}[htdp]
	\begin{center}
\scalebox{0.7}{
	\begin{tabular}{ccccccccccccccc}
\hline
\hline
	Object			&	$M_{ini}$ 	& 	Rotation		&$f_{in}$	&	$M_\mathrm{SN}$	&	$^{4}$He	&	$^{12}$C	&	$^{13}$C	&	$^{14}$N	&	$^{16}$O	&	$^{20}$Ne&	$^{23}$Na&	$^{24}$Mg&	$^{27}$Al	&	$^{28}$Si	\\
\hline
	SMSS 0313-6708	&	60		&	non-rotating	&	0.96	&		37.3			&	1.70e1	&	3.05e-1	&	2.00e-8	&	2.13e-7	&	9.18e-1	&	5.81e-2	&	3.81e-8	&	6.88e-3	&	6.63e-9	&	4.90e-6	\\
					&	80		&	non-rotating	&	0.98	&		46.8			&	2.21e1	&	2.62e-1	&	3.08e-9	&	3.00e-7	&	6.09e-1	&	9.00e-2	&	1.50e-8	&	5.21e-2	&	6.39e-9	&	1.21e-3	\\
	HE 0107-5240		&	30		&	rotating		&	1.07	&		18.1			&	8.21e0	&	4.53e-2	&	4.25e-4	&	2.94e-4	&	6.20e-3	&	4.19e-5	&	3.09e-5	&	2.03e-5	&	2.56e-7	&	3.14e-7	\\
	HE 1327-2326		&	40		&	non-rotating	&	0.96	&		25.5			&	1.09e1	&	1.86e-1	&	2.89e-8	&	1.09e-6	&	5.18e-1	&	2.08e-2	&	1.56e-4	&	1.45e-3	&	4.81e-5	&	5.01e-5	\\
					&	20		&	rotating		&	0.93	&		14.2			&	5.94e0	&	1.59e-1	&	7.78e-3	&	1.18e-2	&	3.36e-1	&	1.83e-2	&	1.77e-4	&	1.20e-3	&	3.81e-5	&	4.11e-5	\\
\hline
	\end{tabular}
}
	\end{center}
	\caption{Stellar yields of the best fit models. The first column shows the object name, from second to fourth columns show the initial mass, the inclusion of rotation, and the adopted $f_{in}$ of the model. The rest show total mass of the ejecta and ejected mass of each element in solar mass units.}
	\label{tab_fit2}
\end{table}

\bibliography{biblio}
\bibliographystyle{apj}


\end{document}